\documentclass{aastex}
\usepackage{natbib}
\usepackage{gensymb}
\usepackage{todonotes}
\usepackage{subfigure}
\usepackage{graphicx}
\usepackage{amsmath}
\usepackage{cases}

\def\yr{\textrm{yr}}

\def\Gpc{\textrm{Gpc}}
\def\kpc{\textrm{kpc}}
\def\pc{\textrm{pc}}

\begin{document}

\title{Astrobiological Effects of Gamma-Ray Bursts in the Milky Way Galaxy}
\slugcomment{Short title: Astrobiological Effects of GRBs in the MW}
\shorttitle{Astrobiological Effects of GRBs in the MW}
\shortauthors{Michael G. Gowanlock}

\author{Michael G. Gowanlock$^1$}
\affil{Massachusetts Institute of Technology, Haystack Observatory, Westford, MA, U.S.A., and Department of Information and Computer Sciences, University of Hawai`i at M\=anoa, Honolulu, HI, U.S.A.}
\email{gowanloc@mit.edu}

\altaffiltext{1}{Massachusetts Institute of Technology, Haystack Observatory, 99 Millstone Road, Westford, MA, 01886, U.S.A. E-mail: gowanloc@mit.edu}

\begin{abstract}
A planet having protective ozone within the collimated beam of a Gamma Ray Burst (GRB) may suffer ozone depletion, potentially causing a mass extinction event to existing life on a planet's surface and oceans.  We model the dangers of long GRBs to planets in the Milky Way and utilize a static statistical model of the Galaxy that matches major observable properties, such as the inside-out star formation history, metallicity evolution, and 3-dimensional stellar number density distribution.  The GRB formation rate is a function of both the star formation history and metallicity; however, the extent to which chemical evolution reduces the GRB rate over time in the Milky Way is still an open question.  Therefore, we compare the damaging effects of GRBs to biospheres in the Milky Way using two models. One model generates GRBs as a function of the inside-out star formation history. The other model follows the star formation history, but generates GRB progenitors as a function of metallicity, thereby favoring  metal-poor host regions of the Galaxy over time. If the GRB rate only follows the star formation history, the majority of the GRBs occur in the inner Galaxy.  However, if GRB progenitors are constrained to low metallicity environments, then GRBs only form in the metal-poor outskirts at recent epochs.  Interestingly, over the past 1 Gyr, the surface density of stars (and their corresponding planets) that survive a GRB is still greatest in the inner galaxy in both models.  The present day danger of long GRBs to life at the solar radius ($R_\odot=8$ kpc) is low. We find that at least $\sim$65\% of stars survive a GRB over the past 1 Gyr. Furthermore, when the GRB rate was expected to have been enhanced at higher redshifts, such as $z\gtrsim0.5$, our results suggest that a large fraction of planets would have survived these lethal GRB events.

\end{abstract}


\section{INTRODUCTION}

With the success of extrasolar planet searches, and subsequent convergence on the fraction of stars that may host Earth-size planets, there is a growing interest in the effects of galactic environments on planetary biospheres~\citep{1995ApJ...444L..53T,2003ApJ...585.1169G,2004Sci...303...59L,2004IJA:240775,2005ApJ...634..509T,2005ApJ...622L.153T,2011AsBio..11..343M,2011AsBio..11..855G,2013RMxAA..49..253C,2013AsBio..13..491J,2013-637X-773-1-6,2014PhRvL.113w1102P,2015AsBio..15..207T,2015ApJ...810L...2D,2015ApJ...810...41L,IJA:10152160,Vukotic2016,2016ApJ...826L...3T,2016A&A...592A..96G}.  One class of potential risks to the habitability of planets are transient radiation events, such as Gamma Ray Bursts (GRBs).  These events are lethal to planets due to the subsequent depletion of ozone in planetary atmospheres.  Furthermore, given that GRBs have collimated emissions that beam radiation on the order of a kpc, they are expected to pose a significant danger to life on planets in general, and Earth in particular. As such, these events may have been responsible for mass extinction events on the Earth~\citep{1995ApJ...444L..53T,2004IJA:240775,2005ApJ...622L.153T,2005ApJ...634..509T,Melott2009,2013-637X-773-1-6,2015AsBio..15..207T}.    

To estimate the effects that GRBs have on the habitability of the Milky Way, determination of the GRB rate from observational and statistical constraints is necessary.  Applying the cosmologically local GRB rate to small volumes is a challenge, as many of the salient characteristics of GRBs, such as their progenitors and environments that give rise to the events are still not well understood.  However, GRBs are found in metal-poor host galaxies~\citep{2006Natur.441..463F,2013ApJ...773..126J}, and thus correlate with low metallicity environments. One explanation for this correlation is the collapsar model of GRB formation, where low metallicity, massive helium stars are the progenitors of long GRBs~\citep{1999ApJ...524..262M}.  Therefore, there is a metallicity bias, where environments that have undergone significant chemical evolution produce fewer GRBs than low metallicity environments.  With the advanced chemical evolution of the Milky Way, it is expected that long GRBs are rare in the Galaxy.  However, the degree to which metallicity quenches GRB formation is still uncertain when applying the locally observed rate to the Galaxy.  

Studies have suggested that the GRB rate is proportional to the star formation rate (SFR)~\citep{1997ApJ...486L..71T,2000MNRAS.312L..35B,2002ApJ...575..111B,2007ApJ...661..394L}; thus, disregarding metallicity bias, the GRB formation history follows the SFR history.  However, this assumption has been challenged, partially due to the environments that GRBs are found~\citep{2006Natur.441..463F}. Despite varying explanations of the GRB formation history in the literature, using the SFR to trace the GRB rate is a reasonable approximation to understanding the GRB formation history in the Milky Way.   Furthermore, this assumption yields a GRB formation history that follows the inside-out formation history of the Milky Way. This is an important effect to capture, as the majority of the stars in the early Galaxy were found within smaller galactocentric radii than at the present day.  Thus, GRBs would have been more lethal to a greater fraction of the overall stars at high redshift than at the present day.

Previous work on the Galactic Habitable Zone~\citep{2011AsBio..11..855G} considered the effects of supernovae on planetary biospheres. An interesting result was that the region with the greatest stellar density (and supernova rate) was found to host the greatest number of habitable planets, at a galactocentric radius of $R\approx2.5~\kpc$.  While the fraction of stars that are nearby a supernova event is much higher at $R\approx2.5~\kpc$ than in the outskirts, or solar neighborhood, the comparatively higher stellar density of the inner Galaxy, and the average age of stars in the region outweigh the negative effects of supernovae.  Another aim of the present study is to observe whether the same phenomenon holds true for GRBs.

 We utilize a model of the Milky Way that considers the inside-out star formation history, chemical evolution, 3-dimensional stellar number density distribution, and determination of the absolute GRB rate of the Galaxy at $z\sim0$. Using these properties, we model the collimated jet emission, and the influence of chemical evolution on GRB formation to examine the effects that long GRBs have on the habitability of the Galaxy. The GRB formation rate is a function of both metallicity and the star formation history (SFH)~\citep{2014ApJS..213...15W}; however, it is unclear to what extent metallicity evolution reduces the GRB rate over time and its influence on the location of GRB progenitors throughout the galactic disk. Therefore we compare two scenarios: (a) where GRB formation is a function of the SFH; and (b) GRB formation follows the SFH where there is a metallicity dependence on GRB progenitors that favor low metallicity host environments.

The paper is outlined as follows. In Section~\ref{sec:methods}, we outline the construction of a model of the Milky Way Galaxy that includes a stellar population consistent with major observable properties. Additionally, we describe the properties of the GRBs, including the beamed emission and two formation history scenarios.  Section~\ref{sec:results} illustrates the results, including two metrics of habitability: 1) the fraction, and 2) the surface density of stars that are within the beam of a GRB over a time period. Additionally, we compare these results to the relevant literature. Finally, we conclude the work in Section~\ref{sec:conclusions}.

\section{METHODS}\label{sec:methods}

\subsection{Stellar Properties}\label{sec:stellarproperties}
In what follows, we outline properties of stars as motivated by observational constraints within the Milky Way. We utilize these properties to generate a model of the stars in the disk of the Milky Way, and to populate distributions of GRBs.  The model assumptions are similar to those utilized in previous work \citep{2011AsBio..11..855G} to populate the stars in the model.  However, we briefly reiterate the model assumptions here as they are also utilized to generate the population of GRBs in the model, which were not considered in previous work.

\subsubsection{The Initial Mass Function}\label{methods:IMF}
We implement the IMF of \citet{2001MNRAS.322..231K}.  The IMF is defined by a two part power law function, where the value $\alpha=1.3$ when $0.08\leq m < 0.5$, and $\alpha=2.3$ when $m \geq 0.5$. In upcoming sections, we demonstrate that this IMF is consistent with the volumetric density of the solar neighborhood when combined with the stellar number density distribution of \citet{2008ApJ...673..864J}. Thus, we do not consider models with other IMFs.

A main sequence lifetime is computed using the equation of \citet{1994sipp.book.....H}:
\begin{equation}T_L=T_{L\odot}\left(\frac{m_\odot}{m}\right)^{2.5},\label{eqn:stellar_lifetime}\end{equation} where $m_\odot=1$ is the Sun's mass, $T_{L\odot}=11$ is the Sun's main sequence lifetime in Gyr \citep{1993ApJ...418..457S}, and $m$ is the star's mass determined from the Kroupa IMF.

\subsubsection{Star Formation History}\label{methods:SFH}
We utilize the SFH reported in Figure 6 of~\citet{2006MNRAS.366..899N}.  The model is consistent with an inside-out formation history of the disk of the Milky Way.  Thus, the early Galaxy is much smaller than at the present day, and GRBs are expected to have an effect on a much larger fraction of the total stars at that epoch.

\subsubsection{Metallicity of the Milky Way}\label{methods:metallicity}
We adopt the chemical evolution model in Figure 11 of~\citet{2006MNRAS.366..899N}, which produces a stellar metallicity, log($Z/Z_{\odot}$), as a function of radial distance, $R$, and time, $t$ (starting at $t=2$ Gyr). As will be discussed below, the GRB rate is determined in part by metallicity; therefore, we use this model of the chemical evolution of the Milky Way to inform the GRB rate.  If a star forms before 2 Gyr, we assign it a metallicity at $t=2$ Gyr.

\subsubsection{Stellar Number Density Distribution}\label{methods:Density}
We utilize the stellar number density distribution of \citet{2008ApJ...673..864J} as follows: 

\begin{equation}\rho_D(R,h)=\rho_D(R,h;L_1,H_1)+f\rho_D(R,h;L_2,H_2),\label{eqn:stellar_density2}\end{equation}
where
\begin{equation}\rho_D(R,h;L,H)=\rho_D(R_\odot,0)e^{R_\odot/L}\times e^{\big(-\frac{R}{L}-\frac{h+h_\odot}{H}\big)}.\label{eqn:stellar_density2a}\end{equation}
$\rho_D$ is the number of stars per pc$^3$, $R$ is the galactocentric distance, $h$ is the vertical height above or below the midplane, $R_\odot=8$ kpc, and $h_\odot=0$ kpc.   From \citet{2008ApJ...673..864J}, we utilize the values $H1$ = 300 pc, $L1$ = 2600 pc, $H2$ = 900 pc, $L2$ = 3600 pc, and $f$ = 0.12 (corresponding to the thin disk scale height and length, the thick disk scale height and length, and the thick-to-thin disk density normalization). To match the total disk mass of $M_*=4.2\times10^{10}M_\odot$ \citep{2008gady.book.....B}, we normalize $\rho_D(R_\odot,h_\odot)=0.084$ stars $\pc^{-3}$.

\citet{2002AJ....124.2721R} find that the stellar density of the solar neighborhood is 0.112 stars pc$^{-3}$. Combining the stellar number density distribution with the Kroupa IMF, the local number density is 76.8\% of that found by \citet{2002AJ....124.2721R}.  Furthermore, the mean mass density of the Milky Way in the model is $3.2\times 10^{-2}$ $M_\odot~\rm{pc}^{-3}$, which is 71\% of the value of  $4.50\pm0.3\times10^{-2}$ $M_\odot~\rm{pc}^{-3}$ reported in~\cite{2001ApJ...554.1274C}.

\subsection{Population of Milky Way Disk Stars}\label{sec:DiskStars}
Utilizing the stellar number density distribution of~\citet{2008ApJ...673..864J} (Section~\ref{methods:Density}), we assign a mass to each star using the \citet{2001MNRAS.322..231K} IMF (Section~\ref{methods:IMF}), which determines the main sequence lifetime of the star (Equation~\ref{eqn:stellar_lifetime}).  Utilizing the radial position of a given star, we assign it a birth date using the inside-out formation history of~\citet{2006MNRAS.366..899N} (Section~\ref{methods:SFH}).  

Our prescription yields a 3-dimensional model of the Galactic disk that matches some of the major observable properties of the Milky Way. We use this distribution of disk stars to record the time(s) they are within the beam of a GRB.  As described in more detail (Section~\ref{method:GRB}), we elect to model the disk of the Galaxy using 3 dimensions, so as to capture the minor effects of GRB beaming orientations on the lethality of GRBs to planets in the Galaxy. We note that we only populate 1\% of the total number of stars that are produced using the parameters above, but we populate 100\% of the GRBs (as described in Section~\ref{method:GRB}).  None of the results are dependent on the choice to model 1\% of the disk stars.  

Unlike previous works that addressed the habitability of the Galaxy, as constrained by supernovae sterilizations \citep{2004Sci...303...59L,2011AsBio..11..855G}, we do not model planet formation as a function of the metallicity gradient of the Milky Way.  While the metallicity gradient has been used to predict planet formation in similar studies~\citep{2004Sci...303...59L,2011AsBio..11..855G}, results from the Kepler mission~\citep{2012Natur.486..375B,2013PNAS..11019273P} show that Earth-mass planets form around stars with a wide range of metallicities, including within low metallicity environments. Therefore, it is sufficient to only model stars, and not attempt to populate Earth-mass planets as a function of metallicity.  Similarly, the work of~\citet{2014PhRvL.113w1102P} examined the fraction of the stars that are affected by GRBs, and did not account for planet formation.  

\subsection{Gamma Ray Bursts in the Milky Way Galaxy}\label{method:GRB}
As a result of recent studies (and increasing convergence) on the GRB rate in the local universe \citep{2010MNRAS.406.1944W,2013ApJ...773..126J,2014MNRAS.444...15H,2015ApJ...812...33S}, there have been new predictions on the frequency of GRBs on smaller volumes, such as the Galaxy \citep{2004ApJ...611.1033F, 2007ApJ...661..394L,2010MNRAS.406.1944W}.  The uncertainty of the GRB rate of the Milky Way is due to both the uncertainty of the progenitors of GRBs, and the influence of metallicity on quenching the GRB formation rate.  For example, \citet{2004ApJ...611.1033F} find that massive Wolf-Rayet stars in binary systems may be GRB progenitors. However, without consensus on GRB progenitors, and estimates of progenitor population size, the GRB rate of the Galaxy has been inferred by utilizing the locally observed GRB rate and the luminosity function of a Milky Way-like galaxy.  The work of \citet{2011AsBio..11..343M} note that the rate of short GRBs is greater than that of long GRBs; however, \cite{2014PhRvL.113w1102P} suggest that the effect of short GRBs on life in the Galaxy is negligible. Therefore, in this work, we focus on the effects on long GRBs.

\citet{2010MNRAS.406.1944W} find that the local GRB rate ($\rho_0$) is $1.3^{+0.6}_{-0.7} \Gpc^{-3}\yr^{-1}$, \citet{2013ApJ...773..126J} calculate that $\rho_0\sim1.3 \Gpc^{-3}\yr^{-1}$, \citet{2014MNRAS.444...15H} show that $0.7 < \rho_0 < 0.8\Gpc^{-3}\yr^{-1}$, and the work of \citet{2015ApJ...812...33S} yields $\rho_0=0.8^{+0.1}_{-0.1} \Gpc^{-3}\yr^{-1}$. Assuming that the rate (uncorrected for metallicity bias) follows the luminosity function \citep{2010MNRAS.406.1944W,2014PhRvL.113w1102P}, then an upper limit on the GRB rate can be obtained for the Milky Way.  This estimate is necessarily an upper limit, as primarily metal-poor host galaxies derive the local GRB rate. Since the Milky Way has undergone significant chemical evolution, and high metallicity environments are likely to quench GRB formation, the GRB rate of the Milky Way is likely significantly lower than that obtained by combining the local rate with the luminosity function of the Milky Way.  The relationship between the effect of metallicity on the GRB rate is not well understood; however, \cite{2013ApJ...773..126J} find that using low metallicity galaxies is required to reproduce the local GRB rate, and \cite{2006Natur.441..463F} show that GRBs are likely associated with low metallicity environments.  An implication of these findings is that GRBs should be rare events in the Milky Way.  \cite{2015ApJS..218...13Y} find that the local GRB rate traces the SFH at redshift $z>1.0$, but not at $z<1.0$; therefore, there are numerous open questions regarding the local GRB rate.

The flux of radiation directed at the atmosphere of a planet determines the degree of lethality imparted by GRBs.  Transient radiation events, such as GRBs (and supernovae), deplete ozone, thus exposing life that may exist on a planet to a potentially lethal flux of radiation from the planet's host star.  Previous works have addressed the effects of supernovae~\citep{2003ApJ...585.1169G,2004Sci...303...59L,2011AsBio..11..855G,2016ApJ...826L...3T} and GRBs~\citep{1995ApJ...444L..53T,2005ApJ...634..509T,2005ApJ...622L.153T,2014PhRvL.113w1102P,2015ApJ...810...41L} on biospheres in the Milky Way.  The work of~\citet{2005ApJ...634..509T} models the evolution of ozone over a period of a month in the Earth's atmosphere after receiving bursts of 10, 100, and 1000 kJ m$^{-2}$, and find that these bursts lead to a depletion of ozone at an altitude of 32 km of -64\%, -91\% and -98\%, respectively. These figures are for local maxima, where the global average level of depletion is -16\%, -36\%, and -65\%, respectively. Ozone depletion can decrease asymptotically, and relatively small fluences can cause significant ozone depletion.  At a distance of 2 kpc from Earth, \cite{2005ApJ...634..509T} find that a typical GRB will deposit a flux of 100 kJ m$^{-2}$, which causes significant damage to the biosphere producing a mass extinction event~\citep{2011AsBio..11..343M}. \citet{2005ApJ...628L..21D} find this distance to be 1 kpc.  While there are varying estimates of the distance at which a typical GRB will threaten life on Earth, we utilize a fixed sterilization distance of 2 kpc. Furthermore, recent results by~\citet{2015AsBio..15..207T} suggest that the UV damage to organisms after O$_3$ depletion may be less than previously found. Therefore, the sterilization distance adopted may overestimate the danger to planets in the Milky Way galaxy, and should be considered a conservative estimate.  The notion of a sterilization event may be better referred to as a mass extinction event; however, we adopt the former term.

As a result of observations of GRBs, and particularly those from \emph{Swift}~\citep{2004ApJ...611.1005G}, advances have been made in determining the structure of GRBs from their luminosity distributions.  In particular, several works have studied the collimated jet structure~\citep{1998ApJ...497..288W,2004ApJ...611.1033F,2007ApJ...661..394L,1999ApJ...522L..39S}.  Two models of the jet structure have emerged: 1) given the opening angle, $\theta_j$, of the jet, there is a uniform energy distribution across the jet and the energy drops outside of $\theta_j$ \citep{1997ApJ...482L..29M,1997ApJ...487L...1R,1998ApJ...499..301M}, and 2) all GRB jets are identical, but with nonuniform energy distributions within the jet~\citep{2002MNRAS.332..945R,2002ApJ...571..876Z,2004ApJ...601L.119Z}, where the energy release is roughly inversely proportional to the squared distance of the axis of $\theta_j$.  In both models, GRBs are collimated, and have a beaming, rather than isotropic emission.

There are different methods that can be used to model the GRB emission. Two methods are as follows: either model the collimated jet emission with the $f_b$ factor and opening angle (described below), or model an isotropic GRB with a lower absolute GRB rate. We elect to model both the jet opening angle, $\theta_j$, and orientation. This model will capture two effects in comparison to the isotropic model. First, in the isotropic GRB case, the GRB will occur at one particular time, whereas when a $\theta_j$ corresponding to the specific $f_b$ is used, then the individual GRBs will occur separately over a time range $\Delta t$.  Secondly, an isotropic GRB will influence all stars once within its radius, whereas modeling the GRBs separately with a specific $f_b$ implies that some stars will be within the beamed emission multiple times, and some will not be within the beam. Electing to model GRBs with a beamed instead of the isotropic emission is likely to have a minor effect on the results. Over a given epoch, this may yield slightly more stars (and associated planets) surviving a GRB, whereas other stars will be within the beam of a GRB multiple times.

We adopt the beaming factor of $f_b^{-1}=50$ from~\cite{2005ApJ...619..412G} that relates the observed number of GRBs to the total number that includes those that are unobservable.  To calculate the opening angle $\theta_j$, we use $f_b=(1-\mathrm{cos}~\theta_j)$, and obtain $\theta_j\approx0.2~\mathrm{rad}$, or a full angle of 22.92$^\circ$. Each GRB is modeled as two 3-dimensional cones (spherical sectors)  that share an apex, where the total length of the axis along $\theta_j$ is 4 kpc (2 kpc in each direction from the apex). The volume of two spherical sectors is as follows: $(4\pi d^3/3)(1-\cos(\theta_j))$, where $d=2$ kpc.

We generate a GRB orientation that yields a uniform distribution in the solid angle. We select two angles for the jet (one spherical sector) as ($\theta_a$, $\phi_a$).  $\theta_a$ is the azimuthal angle, which is uniformly distributed in the range [0, $\pi$], and the inclination angle is $\phi_a=\cos^{-1}(2x-1)$, where $x$ is uniformly distributed in the range [0, 1). The counter-jet spherical sector has an angle ($\theta_b$, $\phi_b$) in the opposite direction that forms the plane along $\theta_j$ through both spherical sectors.

Although the progenitors of GRBs are not well understood, it has been assumed in many studies that the GRB rate is proportional to the SFR~\citep{1997ApJ...486L..71T,2000MNRAS.312L..35B,2007ApJ...661..394L}. For example, \citet{2002ApJ...575..111B} use the SFR to predict the distribution of GRBs as a function of redshift. However, uncertainties in applying the SFR to predict the GRB formation history have been discussed, particularly at high redshift~\citep{2005MNRAS.364L...8N,2011MNRAS.417.3025V}, where both the SFR and lower average metallicities favor a higher GRB formation rate than at low redshift. Several studies have estimated the present day GRB rate of the Milky Way.  \citet{2004ApJ...611.1033F} report a GRB rate of $5\times10^{-5}\yr^{-1}$, \citet{2007ApJ...661..394L} find a GRB rate of $1.67\times 10^{-6} \yr^{-1}$, \citet{2002ApJ566_723S} calculate a GRB rate of $\sim1.0-5.0\times 10^{-7}\yr^{-1}$ (uncorrected for collimation), and \citet{2010MNRAS.406.1944W} show that the GRB rate is $5\times10^{-6} \yr^{-1}$.

We assume that the absolute GRB rate follows both the SFH and the chemical evolution of the Milky Way. We generate a population of GRBs by first generating a mass profile of the Galaxy as a function of $R$, and $t$ (following the inside-out formation history of~\citet{2006MNRAS.366..899N} and the stellar number density distribution of~\citet{2008ApJ...673..864J}, Sections~\ref{methods:SFH}~and~\ref{methods:Density}, respectively). We bin the mass in both temporal and spatial dimensions.  Then we generate a metallicity, log($Z/Z_\odot$), as a function of $R$ and $t$ from~\citet{2006MNRAS.366..899N}, as described in Section~\ref{methods:metallicity}.  Using the absolute GRB rate in the Milky Way at $z\sim0$, we assign a probability that a given mass at $t$ yields a GRB (described below).  This method generates a model where the GRBs are populated following the 3-dimensional stellar number density distribution, SFH, and chemical evolution of the Milky Way.  For example, at early epochs, where star formation is limited to the inner Galaxy, the GRBs and stars (Section~\ref{sec:DiskStars}) will only be populated in this region.

We determine the present day GRB rate (at $z\sim0$) of the Milky Way. The rate, $\rho_{MW}$ is calculated as follows:
\begin{equation}\rho_{MW}=f_b^{-1} \rho_0 V(M_*) f_{Fe},\label{eqn:grbrate}\end{equation}
where $f_b^{-1}$ is the beaming factor, $\rho_0$ is the cosmologically local GRB rate, $V(M_*)$ is the cosmological volume occupied by the Galaxy, where $M_*$ is the total stellar mass of the Galaxy, and $f_{Fe}$ is a metallicity correction factor to account for the metallicity bias of the Galaxy.  We set $f_b^{-1}=50$ (described above from~\cite{2005ApJ...619..412G}). Similarly to~\cite{2015ApJ...810...41L},  for GRBs with $>10^{50}$ erg s$^{-1}$ we adopt $\rho_0=1.6~\rm{Gpc}^{-3} \rm{yr}^{-1}$~\citep{2015ApJ...812...33S}. From~\cite{2015ApJ...810...41L}, $V(M_*)$ is calculated as $V(M_*)=M_*/\rho_*(z)$, where $\rho_*(z)$ is the average stellar density as a function of $z$.  Using $\rho_*(z)=10^{17.46-0.39z} M_\odot~\rm{Gpc}^{-3}$ reported in \cite{2015ApJ...810...41L}, and $M_*=4.2\times10^{10}$ \citep{2008gady.book.....B},  at $z=0$ we obtain $V(M_*)=1.456\times10^{-7}~\rm{Gpc}^{3}$ (roughly one galaxy per 100 Mpc$^{3}$). See \cite{2013ApJ...777...18M} and \cite{2015MNRAS.447....2M} for more information on the stellar mass evolution functions.  The correction factor $f_{Fe}$ accounts for the metallicity bias of the Milky Way.  We adopt $f_{Fe}=0.1$.  This is similar to \cite{2014PhRvL.113w1102P}, as they account for the metallicity of the Milky Way by considering 10\% of the metallicity uncorrected GRB rate.  This is consistent with the notion that the Milky Way has undergone significantly more chemical evolution on average than other galaxies in the local universe.     Other studies suggest that the Milky Way has undergone significant chemical evolution in comparison to other local galaxies~\citep{2006Natur.441..463F,2006AcA....56..333S,2013ApJ...773..126J}, which should suppress the GRB rate, consistent with our selection of $f_{Fe}$.  From Equation~\ref{eqn:grbrate} we obtain $\rho_{MW}=1.165\times10^{-6}$ yr$^{-1}$.

The dependence on $f_{Fe}$ may significantly change the present day GRB rate. \citet{2010MNRAS.406.1944W} calculate a metallicity uncorrected Milky Way GRB rate of $5\times10^{-6}$ yr$^{-1}$. In comparison, the metallicity uncorrected rate is $1.165\times10^{-5}$ yr$^{-1}$ ($f_{Fe}=1$) in our work. Our metallicity corrected rate is $23.3\%$ of the \citet{2010MNRAS.406.1944W} uncorrected rate. This suggests that a 10\% metallicity bias ($f_{Fe}=0.1$) is reasonable for the Milky Way, as it yields a significant fraction of the uncorrected rate in \citet{2010MNRAS.406.1944W}. Furthermore, $f_{Fe}=0.1$ may overestimate the GRB rate of the Galaxy.  \cite{2013ApJ...773..126J} find that in their sample of Milky Way like galaxies from SDSS~\citep{2008MNRAS.391.1117P}, $Z/Z_\odot=0.1$ was found to be an upper limit on GRB host galaxies.  However, they find that only 2\% of the Milky Way has a metallicity below that value. Thus, $f_{Fe}=0.1$ may overestimate the GRB rate of the Milky Way.

The stars are populated in the model as a function of $R$, $t$, and the corresponding metallicity, log($Z/Z_\odot$), for each star is obtained.  The absolute GRB rate is influenced by metallicity; however, it does not explicitly establish that stars with low metallicity should be GRB progenitors. Therefore, we propose two models relating to the location of GRBs throughout the disk of the Milky Way.

\textbf{Model 1} populates GRBs as a function of the SFH. To obtain $\rho_{MW}=1.165\times10^{-6}$~yr$^{-1}$ at $z\sim0$, the probability of a GRB occurring is normalized to the stellar mass produced. We calculate the probability of forming a GRB as a function of mass as: $P_{GRB1}=3.70\times10^{-7}~\mathrm{GRBs}~M^{-1}_\odot$.  Since the majority of the mass in the Galaxy is located at lower galactocentric radii, the majority of the GRBs will occur in that region.   

\textbf{Model 2} populates GRBs as a function of the SFH and the metallicity required of a GRB progenitor. This model yields a GRB rate that evolves as a function of the chemical evolution of the Galaxy. At earlier epochs, the GRB rate should be higher due to a larger fraction of metal-poor stars in the Milky Way.  We populate GRBs as a function of metallicity where lower metallicities enhance the GRB rate.  \citet{2011MNRAS.417.3025V} find that GRBs are produced with metallicities of $\rm{[Fe/H]}=-0.43$, or $0.4Z_\odot$.  We assume that the full GRB rate (Equation~\ref{eqn:grbrate} with $f_{Fe}=1$) is obtained at sufficiently low metallicities, where $Z/Z_\odot\leq0.4$ (or log($Z/Z_\odot)\leq-0.3979$). Furthermore, we assume that the present day rate, $\rho_{MW}$ (with $f_{Fe}=0.1$), is normalized to the average metallicity at the present day in the model which is log($Z/Z_\odot$)=0.0754.  Thus, we obtain a relationship between metallicity and the probability of forming a GRB.  The probability of generating a GRB as a function of metallicity is as follows, where $M_0$ is described below.

{\footnotesize
\begin{subnumcases}{P_{GRB2}(\mathrm{log}(Z/Z_\odot))=}
M_0 & \text{if $\mathrm{log}(Z/Z_\odot)\leq-0.3979$}\label{eqn:metallicity1}\\
M_0[-1.1398\times10^{\mathrm{log}(Z/Z_\odot)}+1.4556] & \text{if $-0.3979<\mathrm{log}(Z/Z_\odot)\leq0.1063$}\label{eqn:metallicity2}\\
0 & \text{if $\mathrm{log}(Z/Z_\odot)>0.1063$}\label{eqn:metallicity3}   
\end{subnumcases}
}

The value of $\mathrm{log}(Z/Z_\odot)=0.1063$ gives $P_{GRB2}=0$ in Equation~\ref{eqn:metallicity2}; therefore, we set the probability to 0 above this metallicity value as shown in Equation~\ref{eqn:metallicity3}.  To obtain the present day GRB rate of $\rho_{MW}=1.165\times10^{-6}$~yr$^{-1}$, the mass normalization is calculated as $M_0=3.106\times10^{-6}~\mathrm{GRBs}~M^{-1}_\odot$. The linear dependence between metallicity abundance and the probability of forming a GRB is utilized as it addresses the notion that at high redshift the metallicity of the Galaxy will be insufficient to quench GRB formation and at low redshift, sufficient chemical evolution will quench GRBs in regions with sufficiently high metallicity.

To match the value of $\rho_{MW}$, the mass normalized probability of forming a GRB in Model~1, $P_{GRB1}$, is 12\% of the value of $M_0$ in Model~2.  Had we elected to substitute the value of $P_{GRB1}$ from Model~1 for $M_0$ in Model~2, then we would only obtain 12\% of the present day GRB rate of $\rho_{MW}$.  This would imply that there are fewer GRB progenitors at $z\sim0$ in Model~2 than Model~1.  An alternate model could explore this scenario; however, we elect to normalize both models to the same rate at $z\sim0$.

Using the metallicity dependence in Model~2 contrasts methods between related literature.  \citet{2014PhRvL.113w1102P} bias the GRB rate by examining the overlap in the metallicity of GRB host galaxies and stars in the Milky Way and \citet{2015ApJ...810...41L} use a similar approach that compares the metallicity of the local universe to the Milky Way.  In these works, metal-poor regions enhance the Milky Way GRB rate, but the regions themselves do not host the majority of the GRBs; rather, high luminosity regions are the predominant GRB hosts.  Therefore, modeling the chemical evolution and allowing GRBs to occur as a function of metallicity allows for a more detailed analysis of the locations of GRBs over the history of the Milky Way.  If future studies demonstrate that GRBs should occur primarily in the regions containing the majority of the stellar mass, then Model~1 may be a more accurate representation of the lethality of GRBs within the Galaxy.

This model will populate GRBs as a function of the SFH, but only in metal-poor environments. Therefore, as the chemical evolution increases over time, GRBs will be quenched in regions of the Milky Way with sufficient metallicity.    This method of generating GRBs is based on both metallicity evolution and the SFH. It may underestimate or overestimate the GRB rate in the past as there is no consensus on how metallicity may influence the GRB rate over the cosmological history of the Milky Way.  Despite these uncertainties, we still make a reasonable assumption regarding the present day GRB rate due to the metallicity bias of the Milky Way and explore two possible scenarios for the distribution of GRBs throughout the Galaxy.

\subsection{Models}
To summarize our model assumptions, using a monte carlo approach, we generate a population of GRBs that follow the 3-dimensional distribution of stars, inside-out formation history and chemical evolution of the disk of the Milky Way.  We assume that GRBs may be beamed in any direction by selecting azimuthal and vertical angles that define the axis along $\theta_j$.  We assume a fixed sterilization distance of 2 kpc.  We populate stars within the Galactic disk, and assign each star a birth date and main sequence lifetime, which follow the formation history of the Milky Way.  For each of these stars, we record the times in which they are within the beam of a GRB.  Both models use an absolute GRB rate for the Milky Way, $\rho_{MW}$, where one model populates GRBs as a function of the SFR, and the other uses the SFR, but is constrained  to populating GRBs in low metallicity environments.

\subsection{SFH and Chemical Evolution}

\begin{figure*}[htp]
\centering
      \subfigure[]{
            \includegraphics[width=0.45\textwidth]{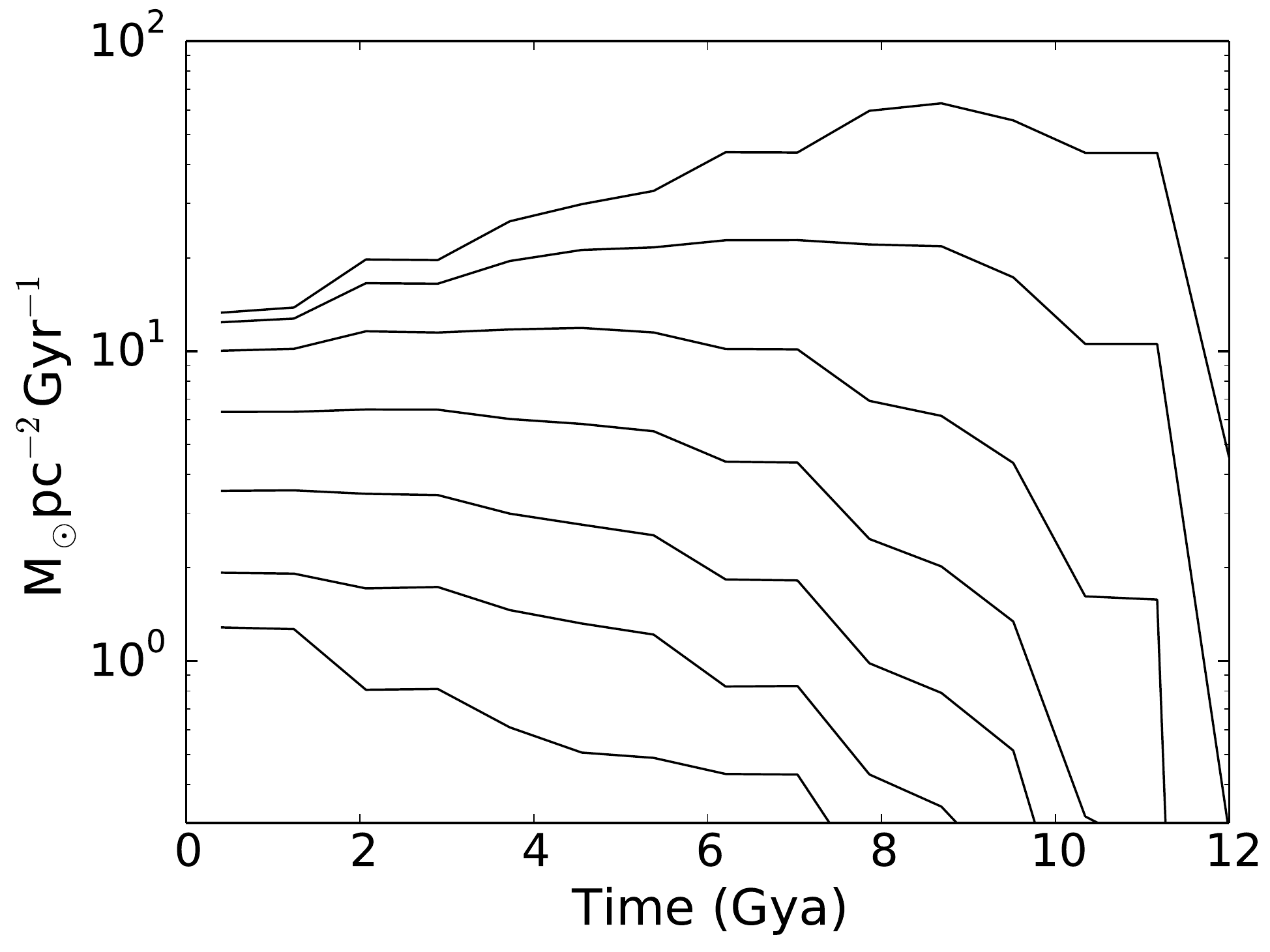}
      }     
        \subfigure[]{
            \includegraphics[width=0.45\textwidth]{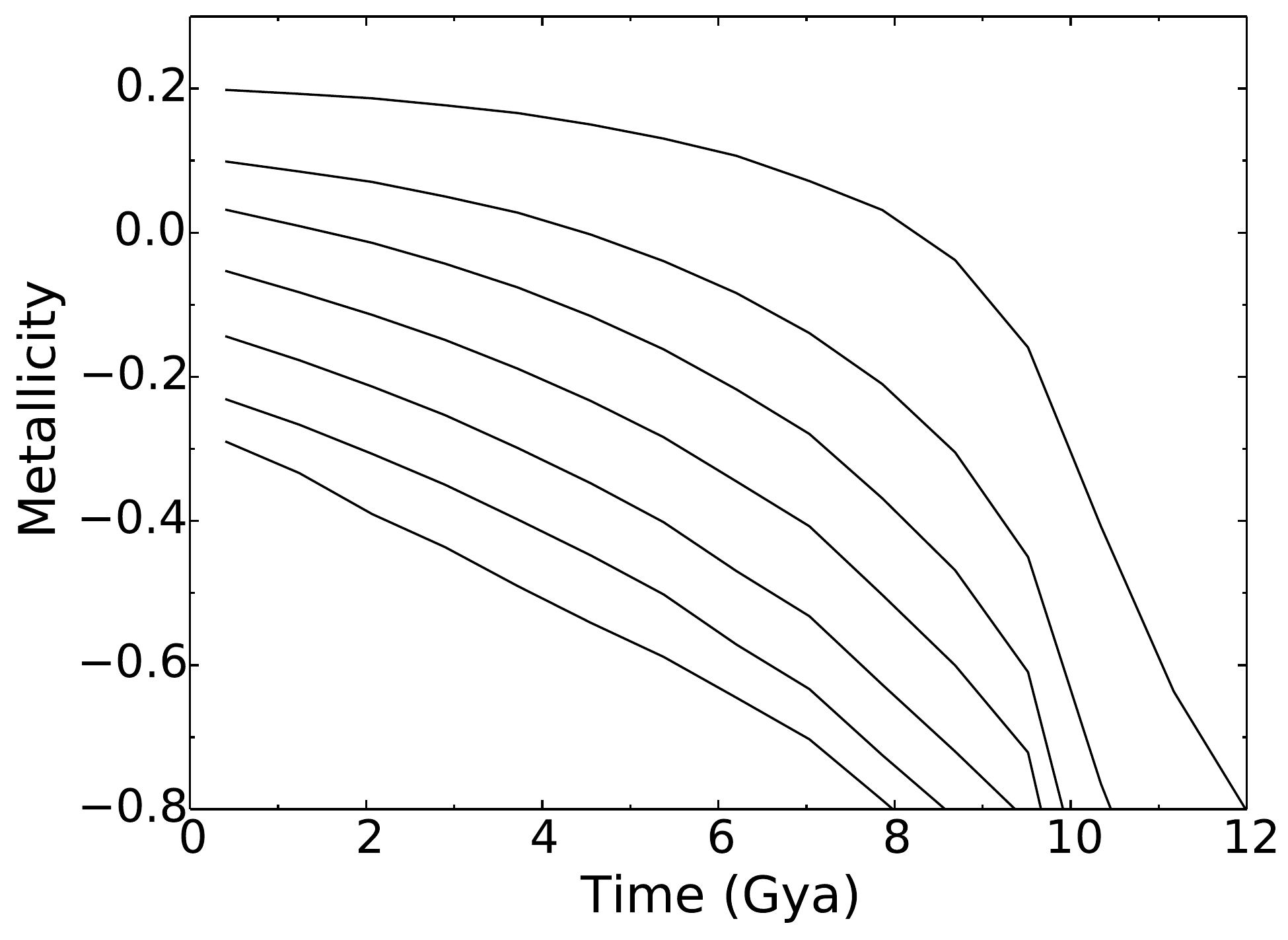}
      }

    \caption{(a) The formation history of the stars populated in the model. (b) The metallicity evolution of the mass populated in the model to generate the GRBs in log($Z/Z_\odot$). The curves correspond to 2 kpc, 4 kpc, 6 kpc, 8 kpc, 10 kpc, 12 kpc, and 14 kpc, ordered from top to bottom.}
   \label{fig:models_formation_history}
\end{figure*}

Figure~\ref{fig:models_formation_history}~(a) plots the formation history of the stars populated in the model in 2 kpc increments from 2 kpc (top curve) to 14 kpc (bottom curve).  We utilize the inside-out formation history of~\citet{2006MNRAS.366..899N}, and note that we populate 1\% of the stars in the distribution.  Figure~\ref{fig:models_formation_history}~(b) plots the chemical evolution of the metallicity.  Both the SFH and chemical evolution are used to determine the absolute GRB rate, $\rho_{MW}$, in the models.  Note that the inner Galaxy undergoes more chemical evolution due to earlier star formation than the outskirts.

\begin{figure*}[htp]
\centering
            \includegraphics[width=0.5\textwidth]{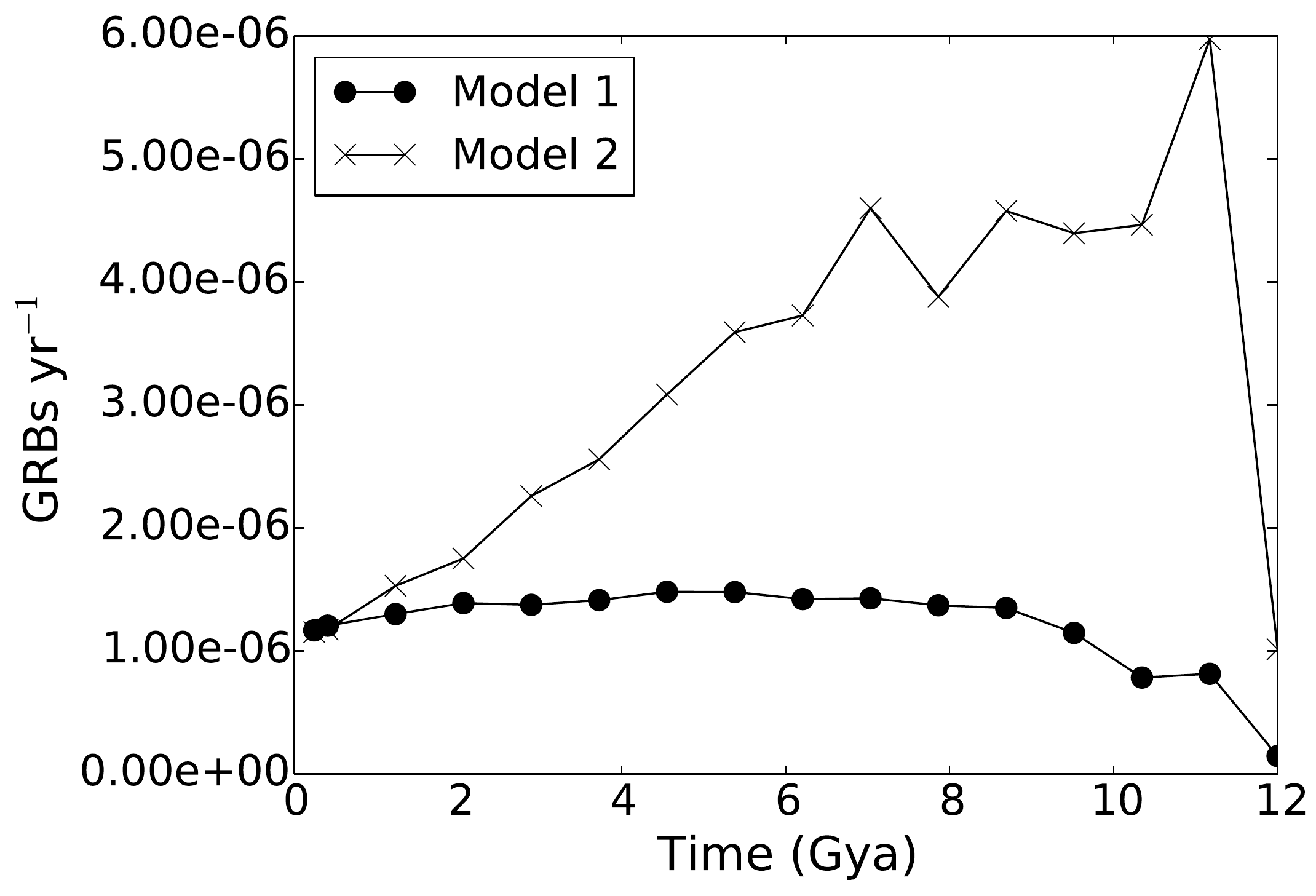}   
    \caption{Comparison of the GRB rates over time in the two models. GRBs are populated as a function of the SFR in both models; however, GRBs in Model~2 are only populated in metal-poor environments. The GRB rate is fairly consistent throughout time in Model~1, whereas in Model~2 the rate is enhanced at high redshift, and suppressed at the present day.}
   \label{fig:GRB_rates_model1_vs_model2}
\end{figure*}

We compare the GRB formation rate over time to illustrate the difference between the models.  In upcoming sections we will show where the GRBs occur within the disk of the Milky Way. Figure~\ref{fig:GRB_rates_model1_vs_model2} contrasts GRB rates between the two models over time.  From the plot, we see that both models match the present day GRB rate, $\rho_{MW}$. However, we observe that the GRB rate in Model~2 was significantly higher in early epochs of the Milky Way as a result of insufficient chemical evolution to quench GRB formation. Comparatively, the GRB rate in Model~1 at early epochs is much lower than Model~2, as the GRB rate is not enhanced by low metallicity environments in the model.

\section{RESULTS}\label{sec:results}

\subsection{Model~1}
\begin{figure*}[htp]
\centering
      
        \subfigure[]{
            \includegraphics[width=0.45\textwidth]{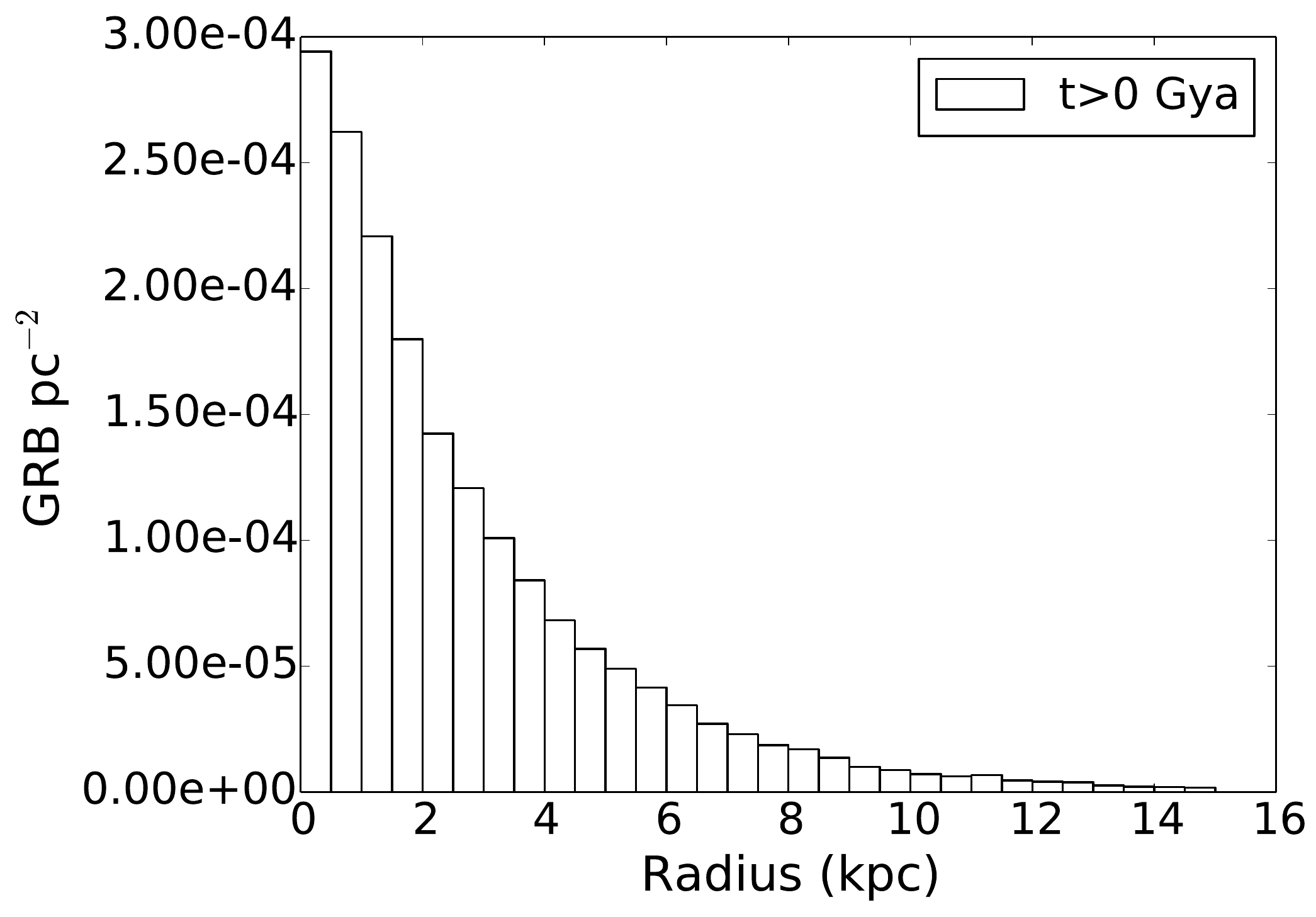}
      }
        \subfigure[]{
            \includegraphics[width=0.45\textwidth]{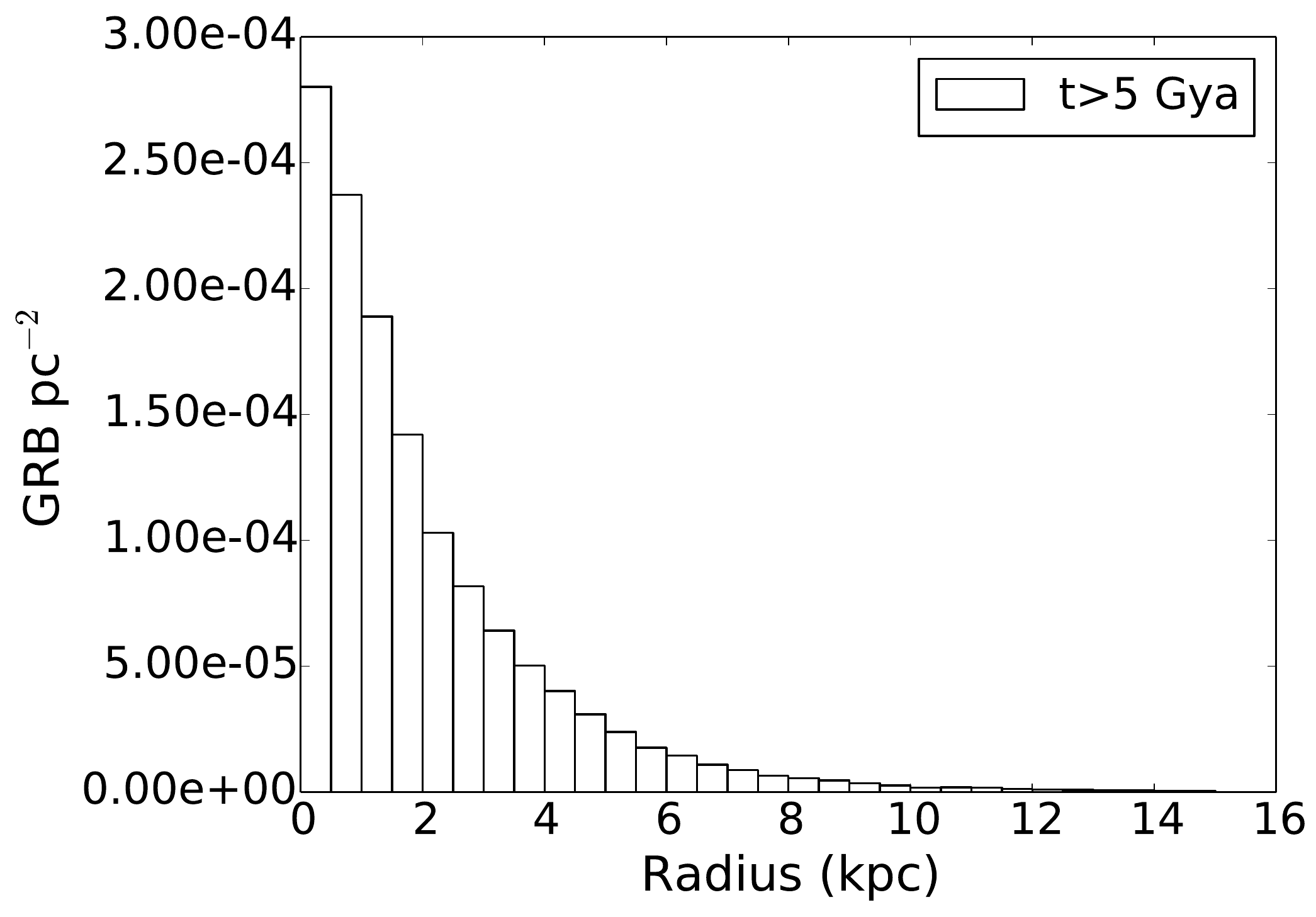}
      }     
        \subfigure[]{
            \includegraphics[width=0.45\textwidth]{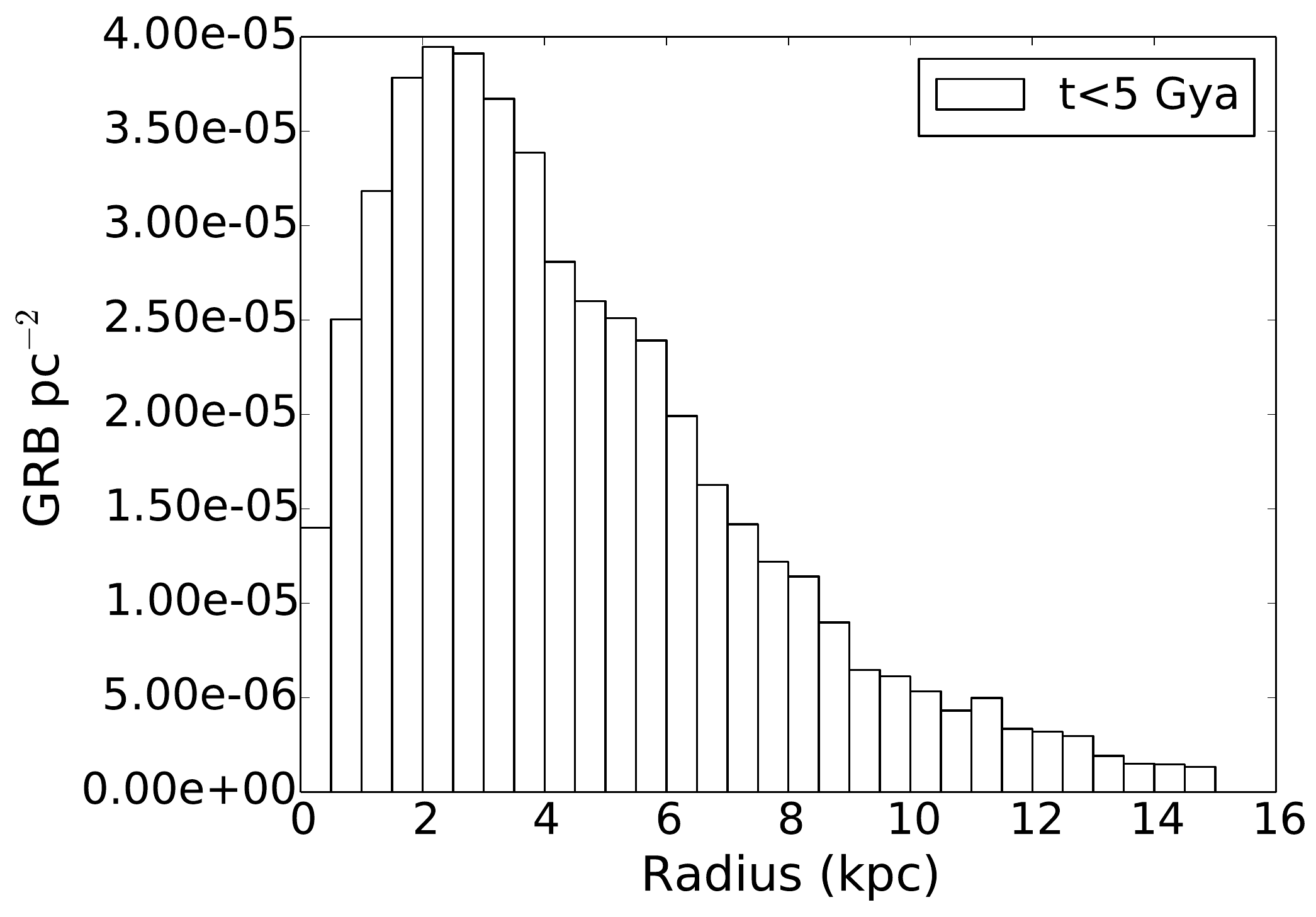}
      }   
         \subfigure[]{
            \includegraphics[width=0.45\textwidth]{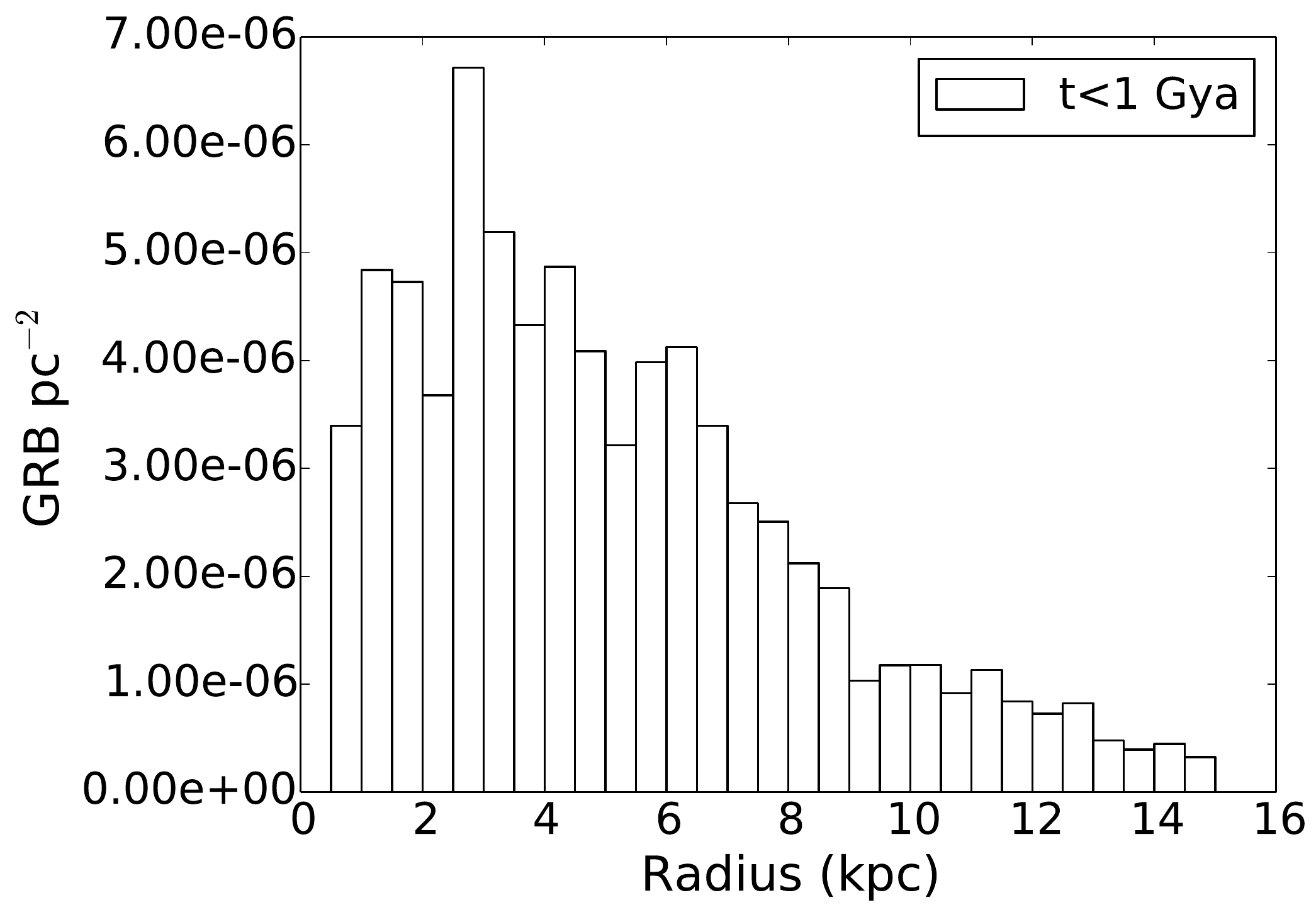}
      }

    \caption{The number of GRBs pc$^{-2}$ populated in Model~1 in 0.5 kpc bins for (a) the entire history of the Milky Way, or $t>0$ Gya, (b) excluding the last 5 Gyr, or $t>5$ Gya, (c) within the last 5 Gyr, $t<5$ Gya, and (d) within the last 1 Gyr, $t<1$ Gya.}
   \label{fig:model1_grb}
\end{figure*}

In Model~1, the locations of GRBs are populated according to the SFH, where the probability of forming a GRB is given by the mass normalized probability, $P_{GRB1}$ (Section~\ref{method:GRB}). Figure~\ref{fig:model1_grb} plots the surface density of GRBs in the model over differing time periods. Figure~\ref{fig:model1_grb}~(a) plots the distribution of GRBs over the entire simulation, Figure~\ref{fig:model1_grb}~(b) excludes the past 5 Gyr, Figure~\ref{fig:model1_grb}~(c) shows those GRBs within the last 5 Gyr, and Figure~\ref{fig:model1_grb}~(d) within the last 1 Gyr.  We plot the location of the source of the GRB, and do not capture the location of the collimated beams. Since the majority of the mass is located at low galactocentric radii, we find that the majority of the GRBs are located in that region over all epochs.

\subsubsection{Sterilization Distributions}\label{results:model1_fraction}

\begin{figure*}[htp]
\centering
	  \subfigure[$R=2.5$ kpc]{
            \includegraphics[width=0.4\textwidth]{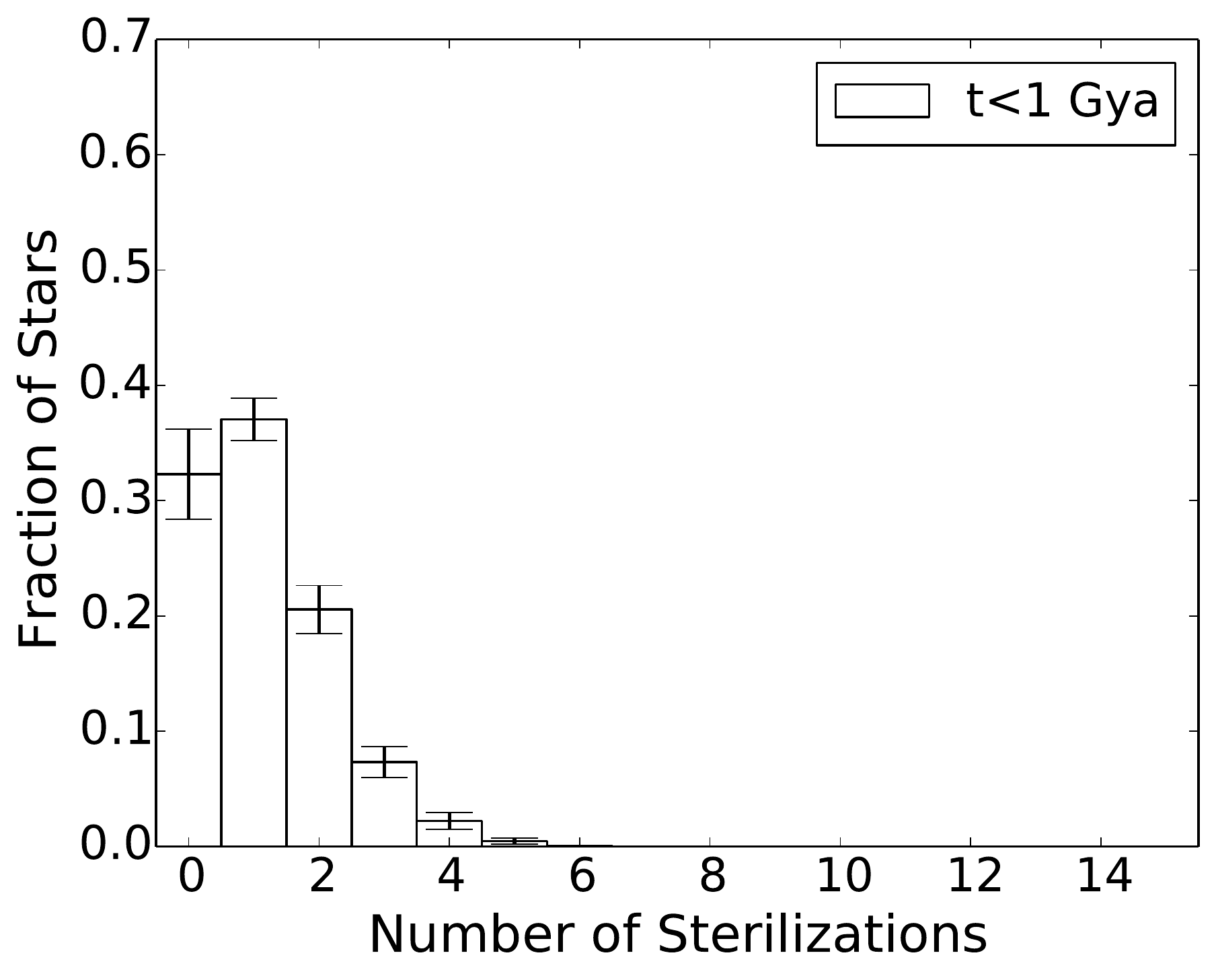}
      }     
         \subfigure[$R=2.5$ kpc]{
            \includegraphics[width=0.4\textwidth]{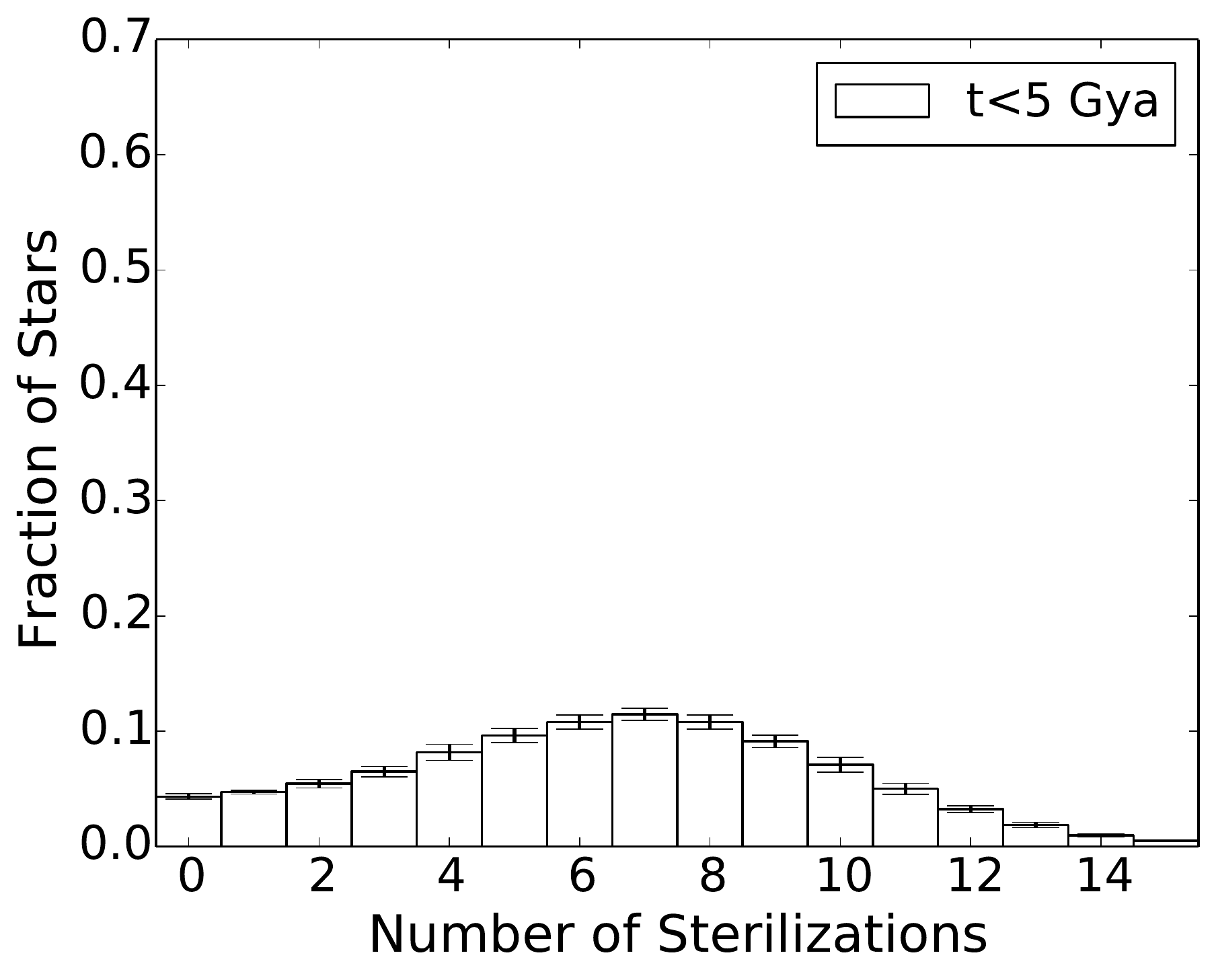}
      } 
      \subfigure[$R=8$ kpc]{
            \includegraphics[width=0.4\textwidth]{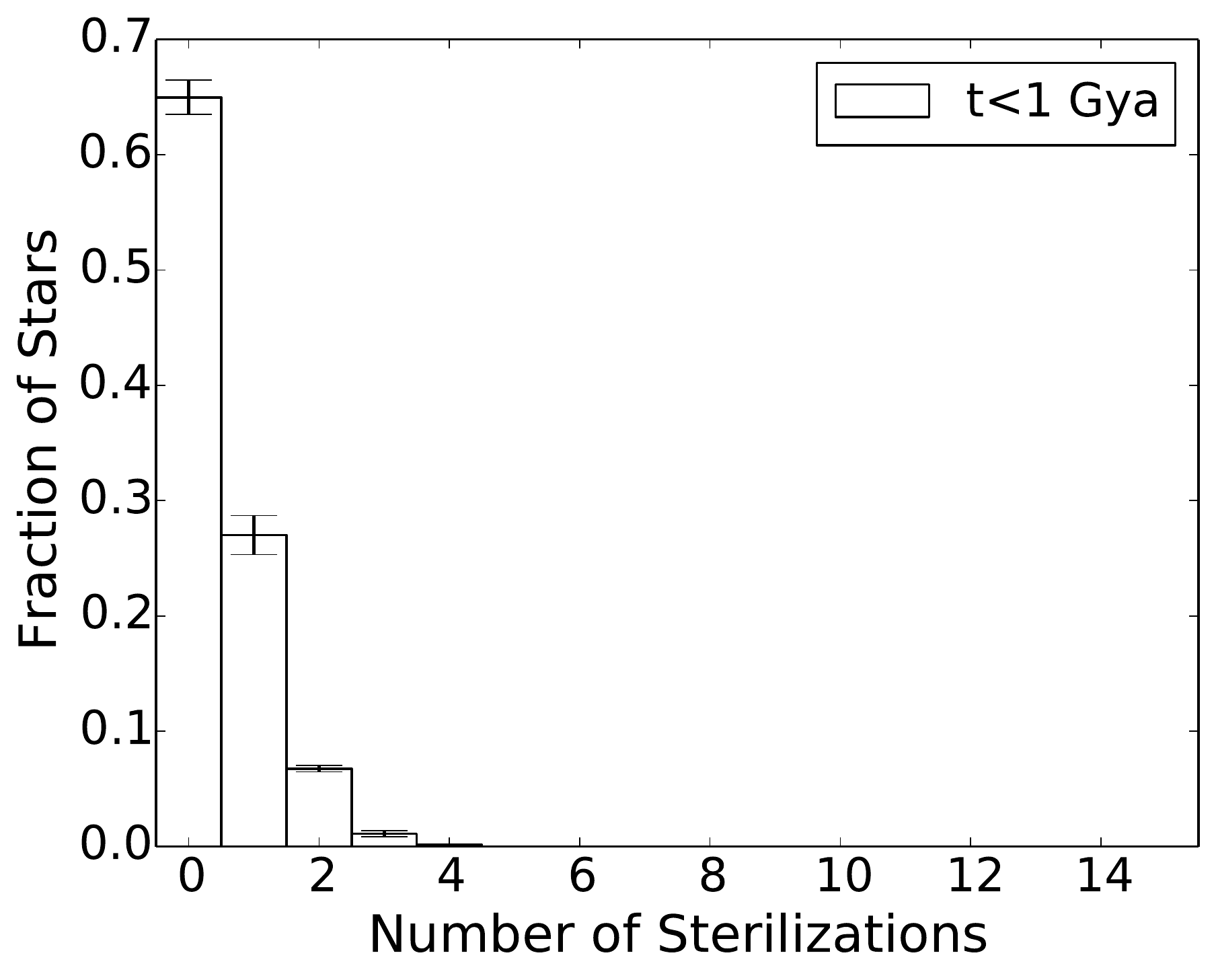}
      }     
         \subfigure[$R=8$ kpc]{
            \includegraphics[width=0.4\textwidth]{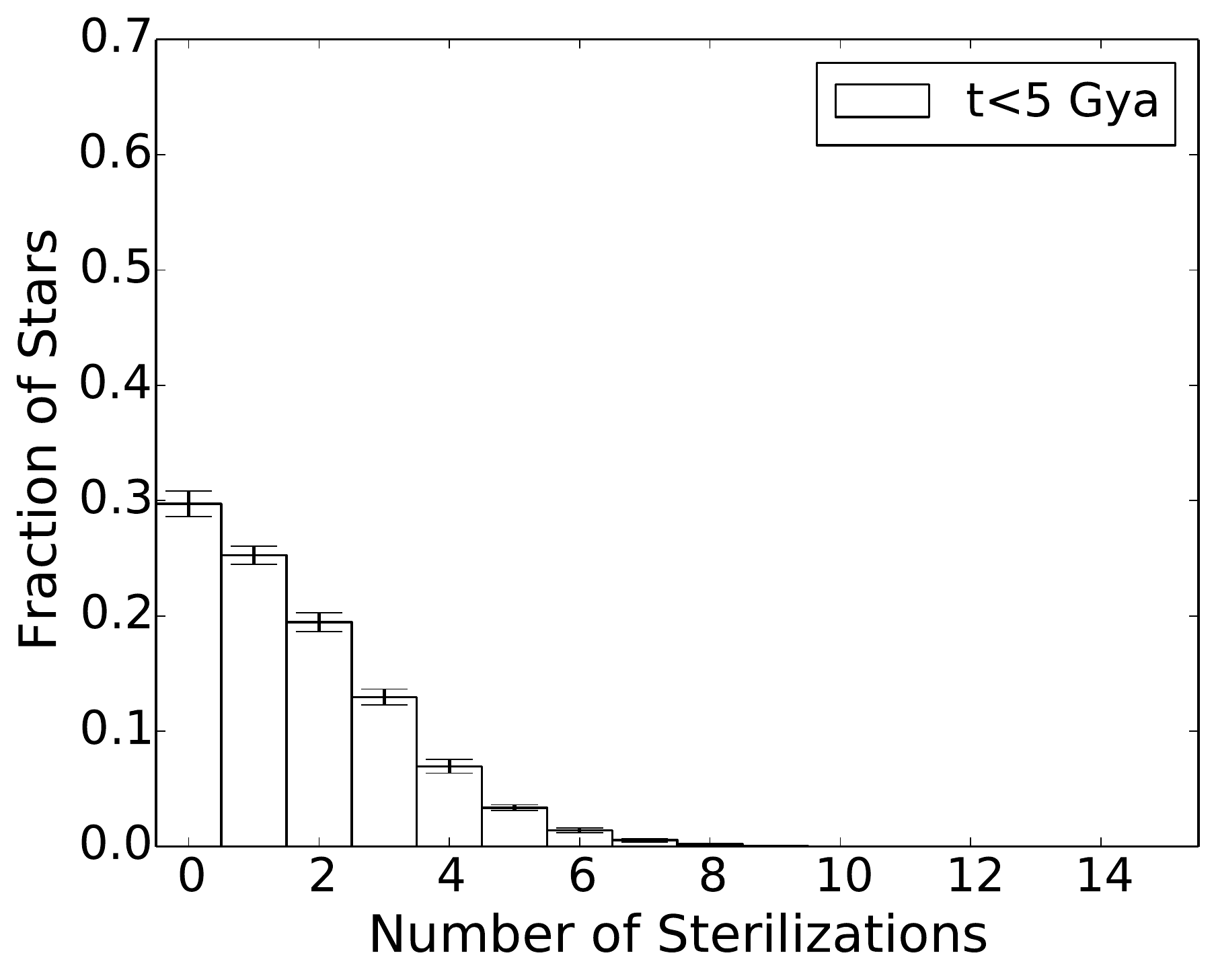}
      }

    \caption{The fraction of stars sterilized by the number of GRBs indicated on the horizontal axis over the past 1 and 5 Gyr at 2.5 kpc (a) and (b) and 8 kpc (c) and (d), respectively. Error bars represent 90\% confidence intervals.}
   \label{fig:last_1Gyr_5_gyr_25_8kpc_Model1}
\end{figure*}

Figure~\ref{fig:last_1Gyr_5_gyr_25_8kpc_Model1} shows the distribution of the number of sterilizations at $t<1$ Gya and $t<5$  Gya at two regions, $R=2.5$ kpc and $R=8$ kpc.  Most stars do not survive GRBs in the inner Galaxy at $R=2.5$ kpc over the past 5 Gyr (Figure~\ref{fig:last_1Gyr_5_gyr_25_8kpc_Model1}~(b)).  Interestingly, we find that at $R=8$ kpc over the past 1 Gyr, there is a $\sim$65\% chance of surviving a GRB.  This is a higher survivability probability than the work of~\citet{2014PhRvL.113w1102P} that find a $\sim40$\% chance of surviving a GRB within the last 1 Gyr.  We also note that at $t<5$ Gya, $\sim30\%$ of the stars survive GRBs at $R=8$ kpc (Figure~\ref{fig:last_1Gyr_5_gyr_25_8kpc_Model1}~(d)). This suggests that there have been numerous planets that have not been within the beam of a GRB since the Earth formed at our galactocentric radius even when ignoring that GRBs should only form in low metallicity environments in the Milky Way (low metallicity GRB progenitors are addressed in Model~2). Therefore, the notion that $\sim30\%$ of planets survive a GRB over the past 5 Gyr may overestimate the lethality of GRBs. Furthermore, an exposure rate of $\sim1$~GRB per 5~Gyr is unlikely to be prohibitive to the long term habitability of a planet.

\subsubsection{Surface Density of Surviving Stars}\label{results:model1_number}

\begin{figure*}[htp]
\centering
            \includegraphics[width=0.5\textwidth]{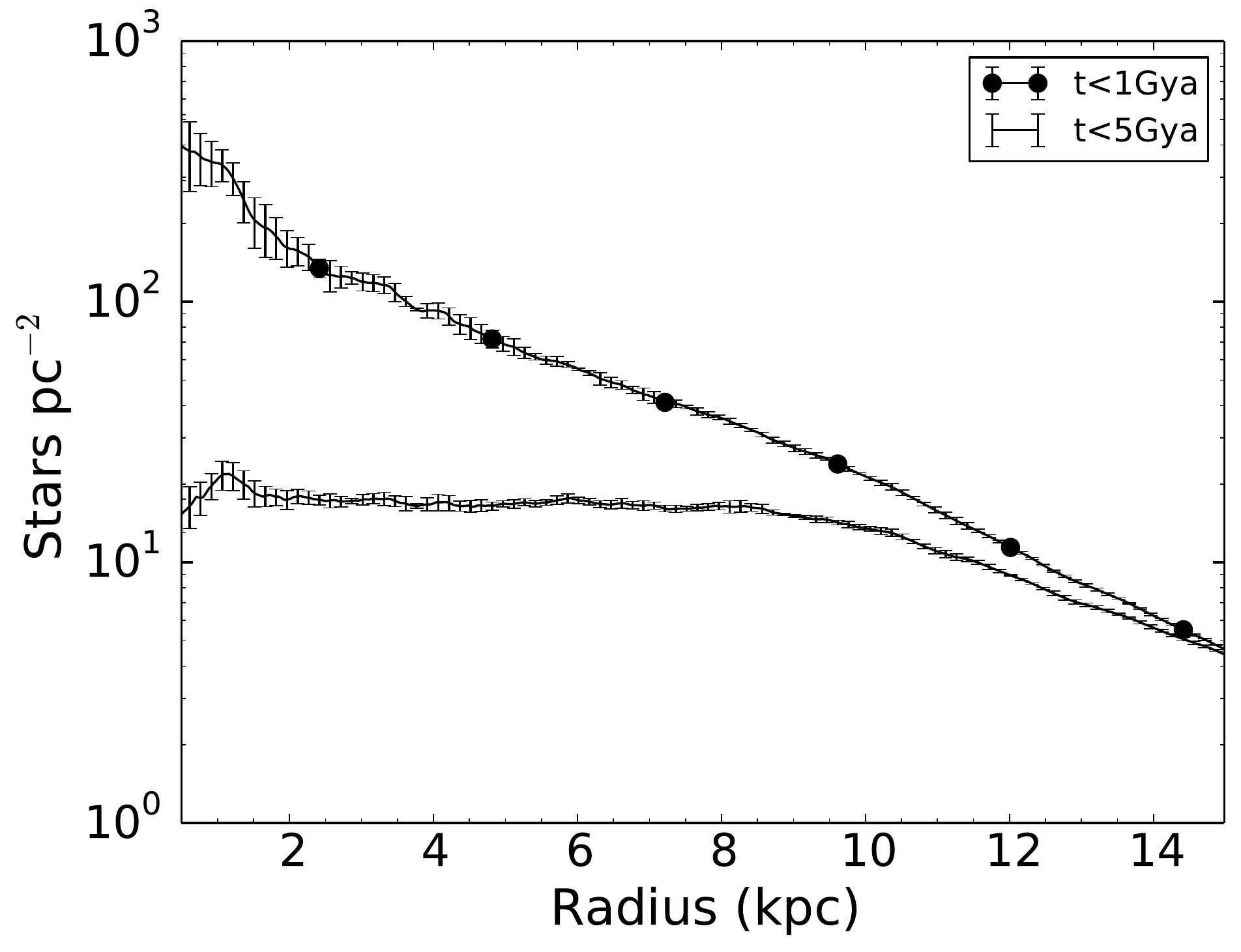}   
    \caption{The surface density of stars that are not within the beam of a GRB over the past 1 Gyr (upper curve) and 5 Gyr (lower curve). Error bars represent 90\% confidence intervals.}
   \label{fig:starsunsterilizedlastonegyrfivegyr-Model1}
\end{figure*}

Beyond examining the fraction of stars that are within the beam of a lethal GRB, we now examine the area density of stars that survive a GRB over the past 1 and 5 Gyr as shown in Figure~\ref{fig:starsunsterilizedlastonegyrfivegyr-Model1}.  Despite the model predominantly hosting GRBs in the inner Galaxy (Figure~\ref{fig:model1_grb}), we find that the surface density of surviving stars is greatest in the inner Galaxy over the past 1 Gyr, and that the surface density of surviving stars is roughly consistent between $2.5\lesssim R\lesssim8$ kpc over the past 5 Gyr. Over the past 1 Gyr, we find that the surface density of surviving stars at $R=2.5$ kpc is $\sim 128$ stars pc$^{-2}$ whereas $\sim 36$ stars pc$^{-2}$ survive at $R=8$ kpc. This is an interesting result as the region of the Galaxy that has the greatest luminosity (the inner Galaxy) hosts the majority of GRBs; however, it also contains the majority of the planets in the Milky Way that survive GRB events.



\subsection{Model~2}
\begin{figure*}[htp]
\centering
      
        \subfigure[]{
            \includegraphics[width=0.45\textwidth]{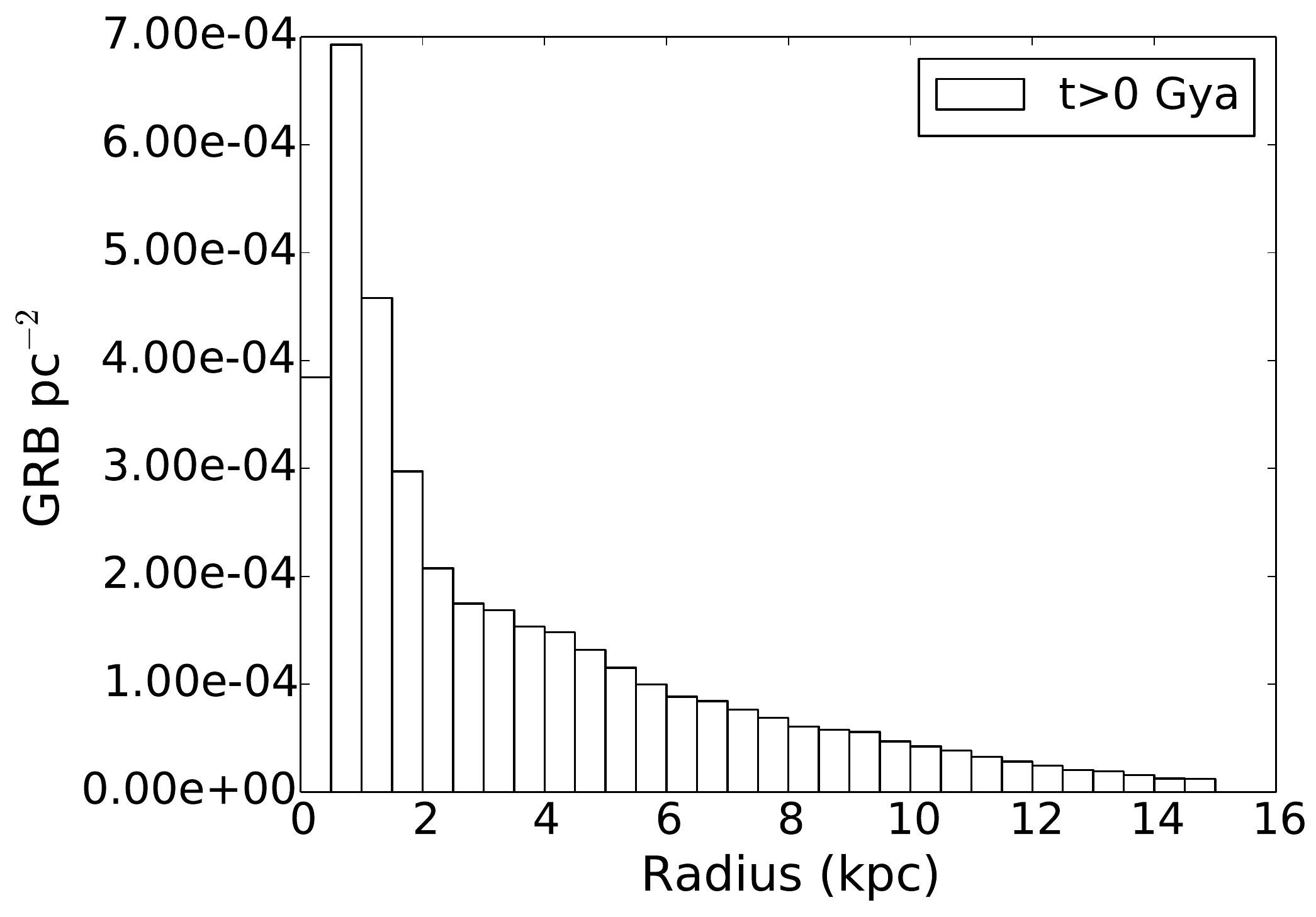}
      }
        \subfigure[]{
            \includegraphics[width=0.45\textwidth]{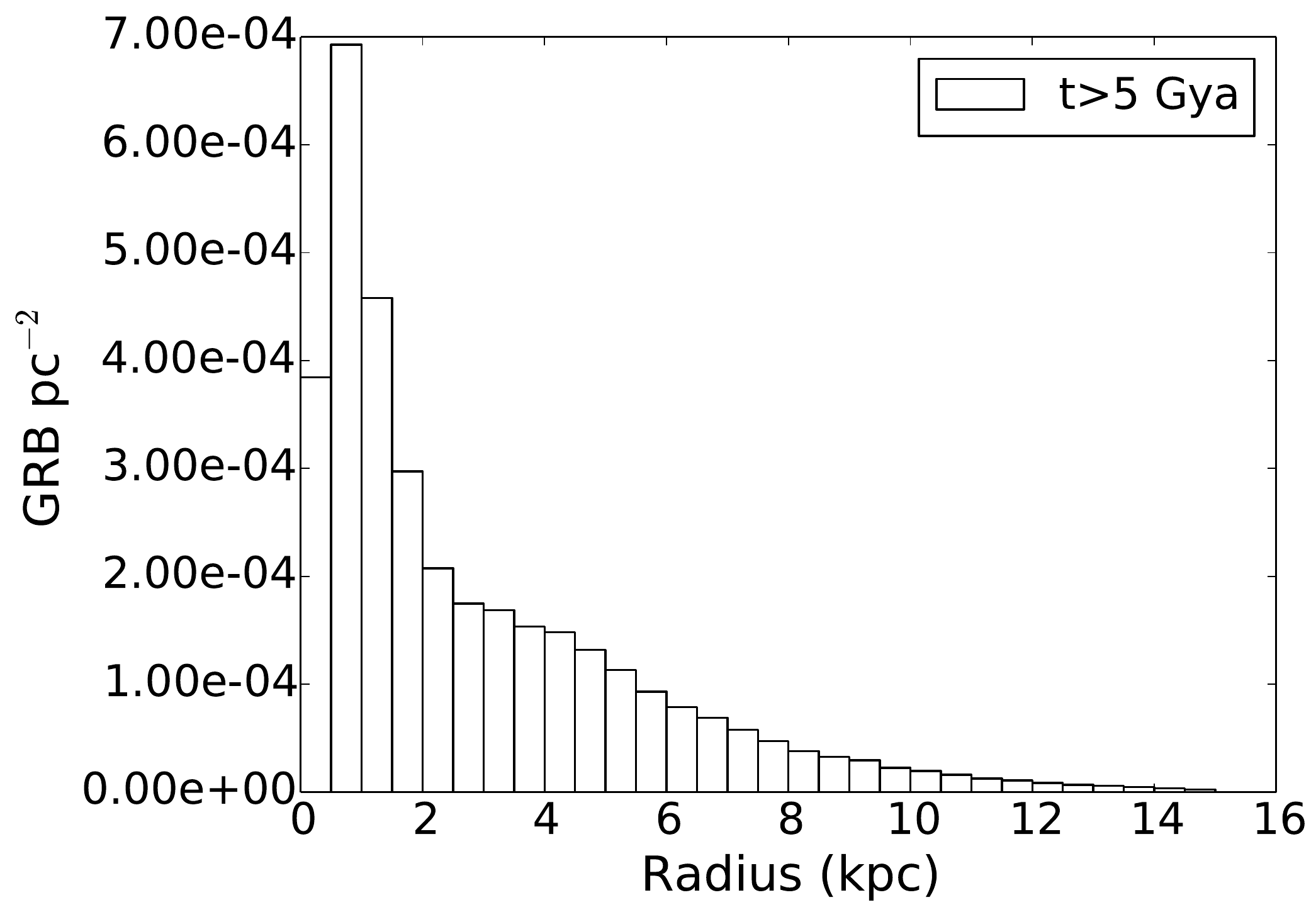}
      }     
        \subfigure[]{
            \includegraphics[width=0.45\textwidth]{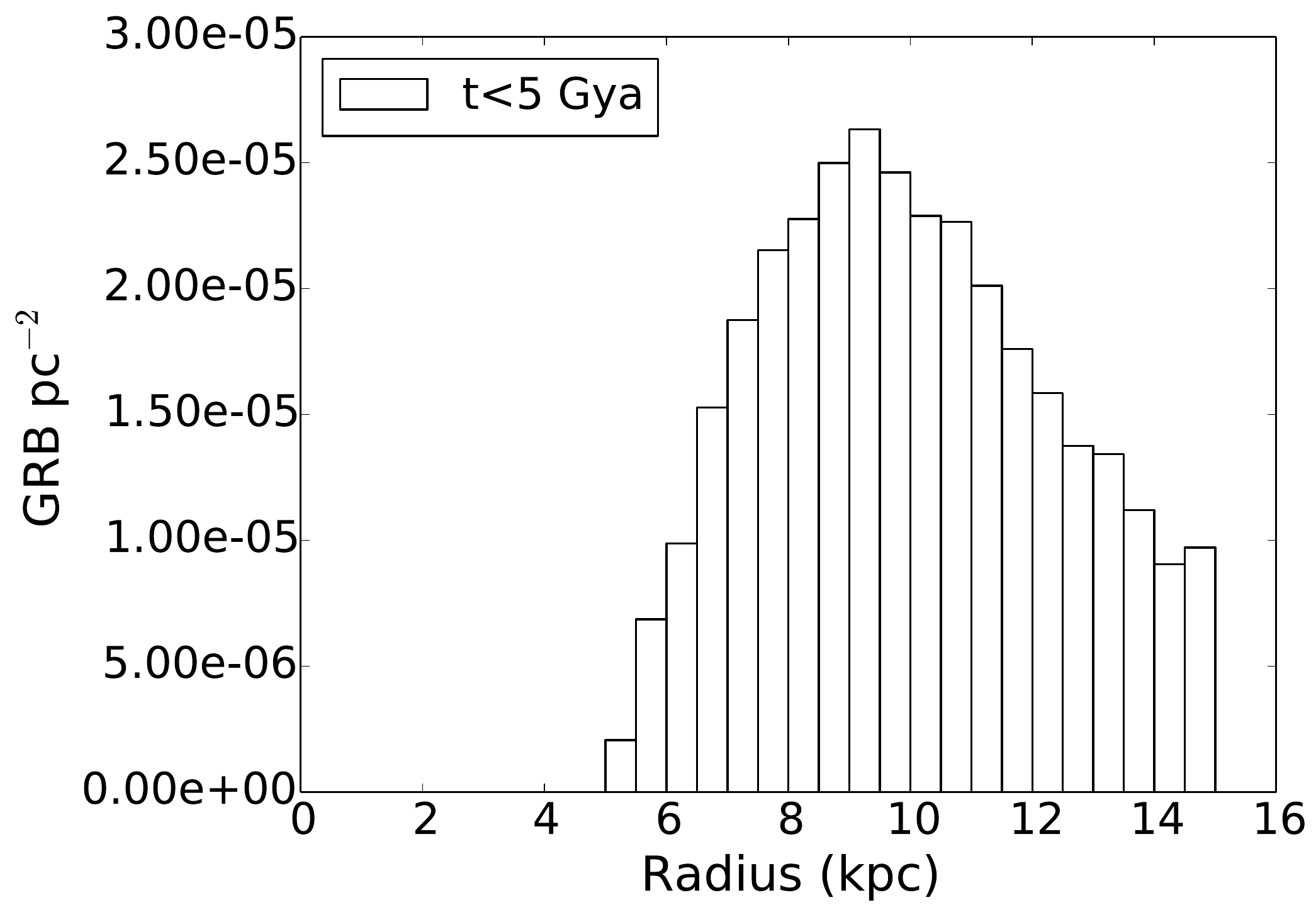}
      }   
         \subfigure[]{
            \includegraphics[width=0.45\textwidth]{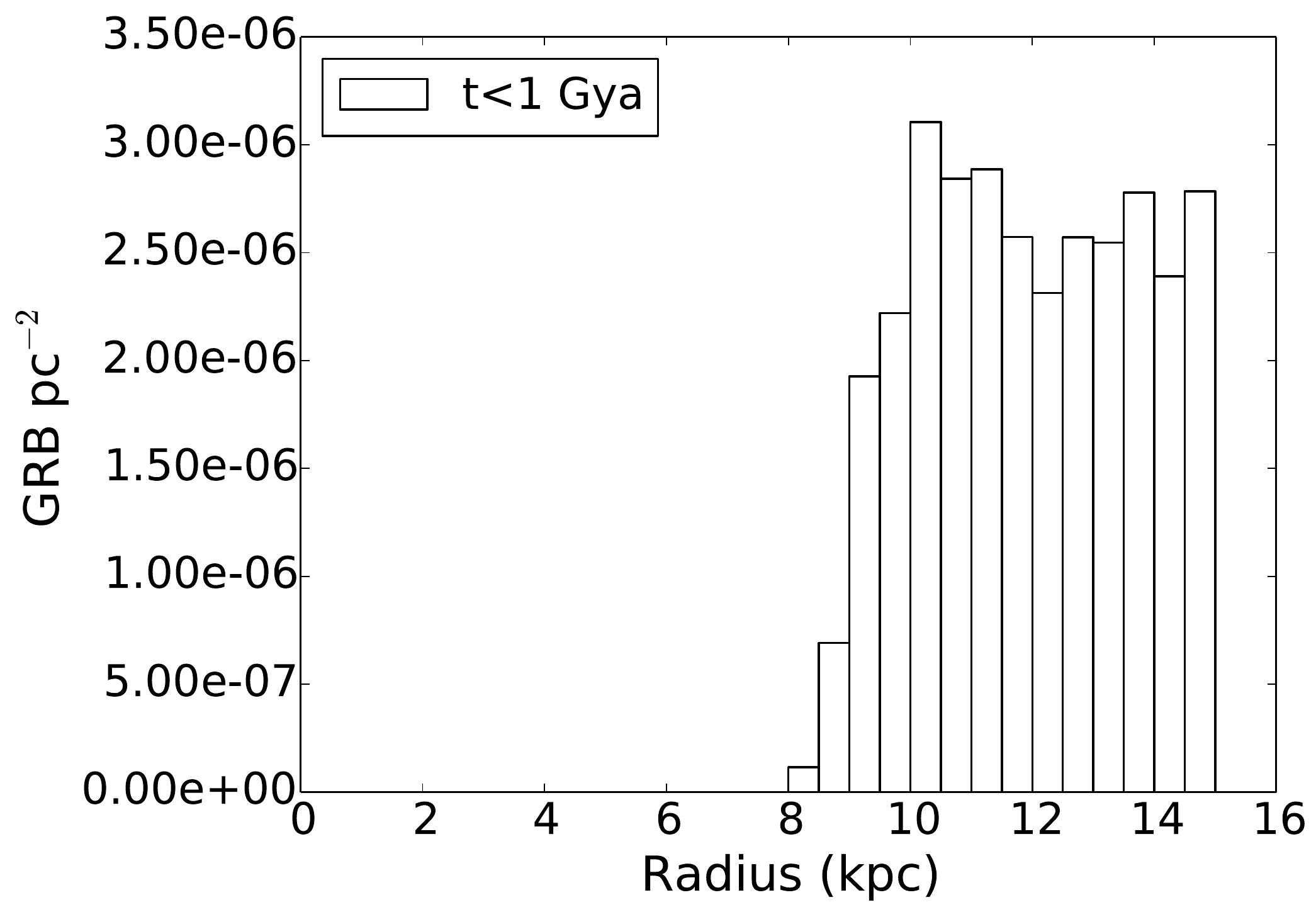}
      }

    \caption{The number of GRBs pc$^{-2}$ populated in Model~2 in 0.5 kpc bins for (a) the entire history of the Milky Way, or $t>0$ Gya, (b) excluding the last 5 Gyr, or $t>5$ Gya, (c) within the last 5 Gyr, $t<5$ Gya, and (d) within the last 1 Gyr, $t<1$ Gya.}
   \label{fig:model2_grb}
\end{figure*}

In Model~2, GRBs are populated according to the SFH and have a metallicity dependence as outlined by $P_{GRB2}$ (Section~\ref{method:GRB}). Figure~\ref{fig:model2_grb} plots the surface density of GRBs in the model over differing time periods similarly to Figure~\ref{fig:model1_grb}. From Figure~\ref{fig:model2_grb}~(a) we observe that the inner Galaxy hosts the majority of the GRBs integrated over all epochs; however, at later epochs, such as within the last 1 Gyr (Figure~\ref{fig:model2_grb}~(d)), the chemical evolution is sufficient such that we do not expect any GRBs to occur within $R\lesssim8$ kpc of the disk of the Milky Way. Therefore, if we assume that GRBs only occur in low metallicity environments, then the Milky Way is likely to host very few GRBs at recent epochs ($z\lesssim 0.5$) that are within regions with high stellar densities.  Thus, late epoch GRBs are likely to sterilize a small fraction of the total number of planets within the Milky Way.

\subsubsection{Sterilization Distributions}\label{results:model2_fraction}

\begin{figure*}[htp]
\centering
      \subfigure[$R=8$ kpc]{
            \includegraphics[width=0.4\textwidth]{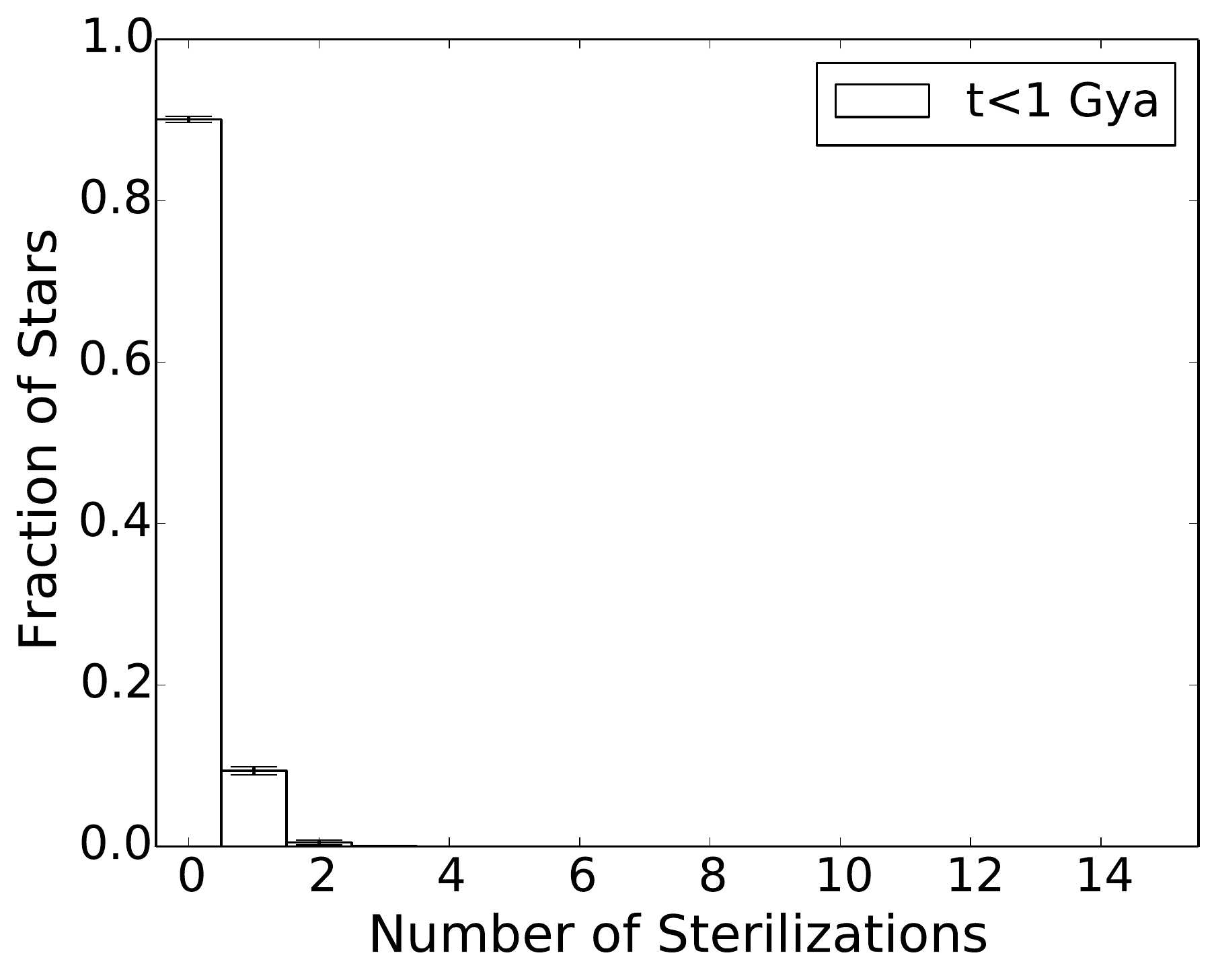}
      }     
         \subfigure[$R=8$ kpc]{
            \includegraphics[width=0.4\textwidth]{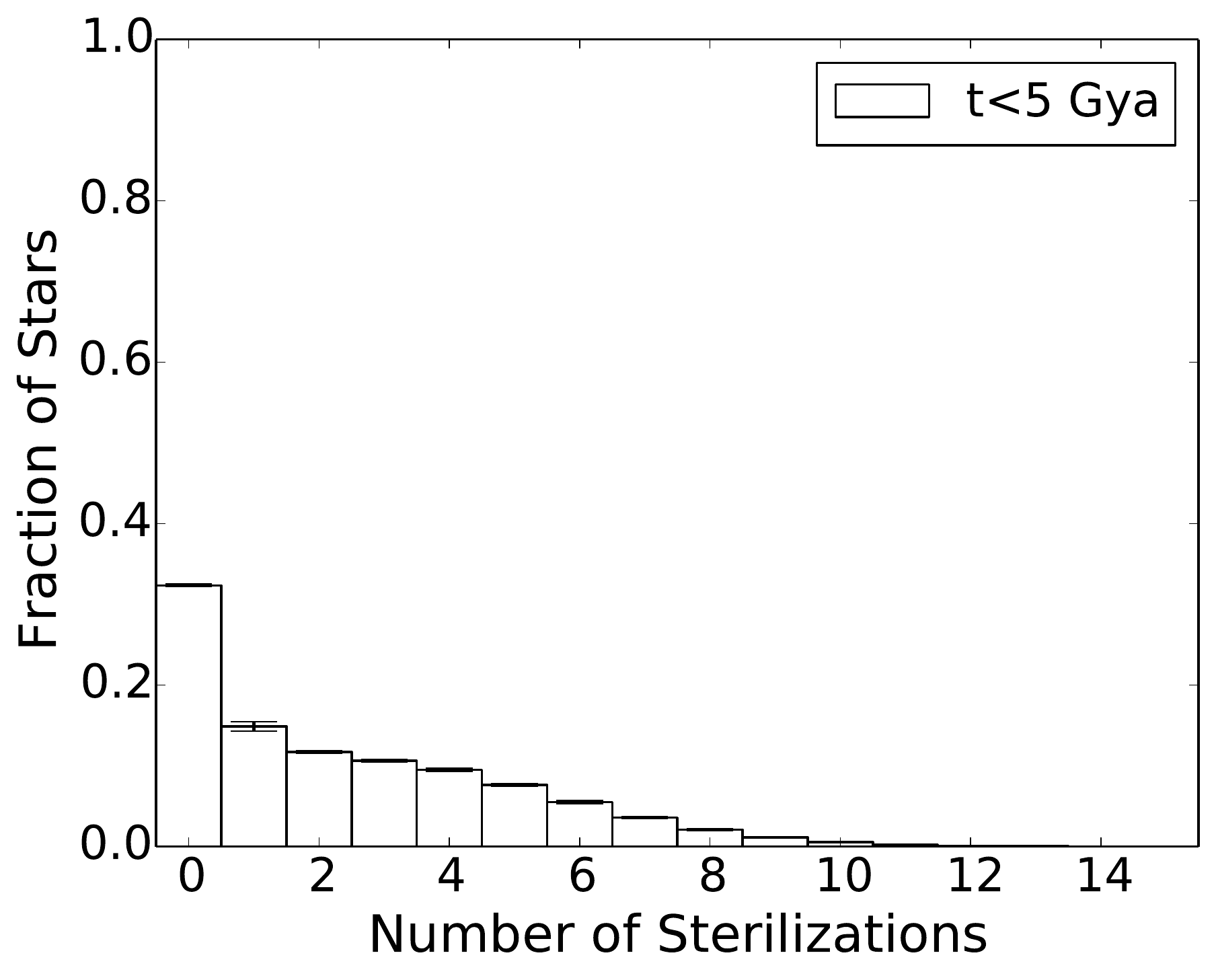}
      }
      \subfigure[$R=14$ kpc]{
            \includegraphics[width=0.4\textwidth]{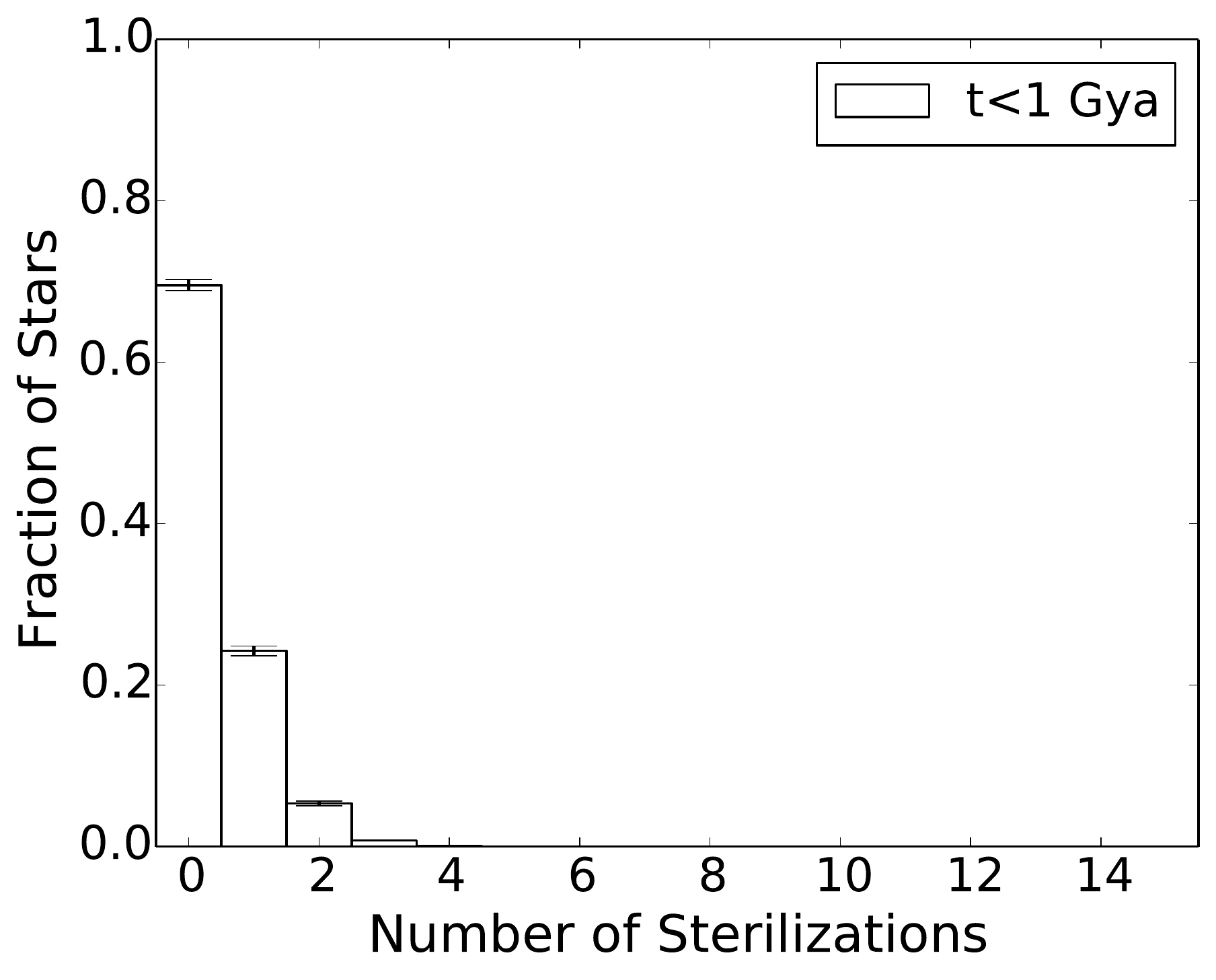}
      }     
         \subfigure[$R=14$ kpc]{
            \includegraphics[width=0.4\textwidth]{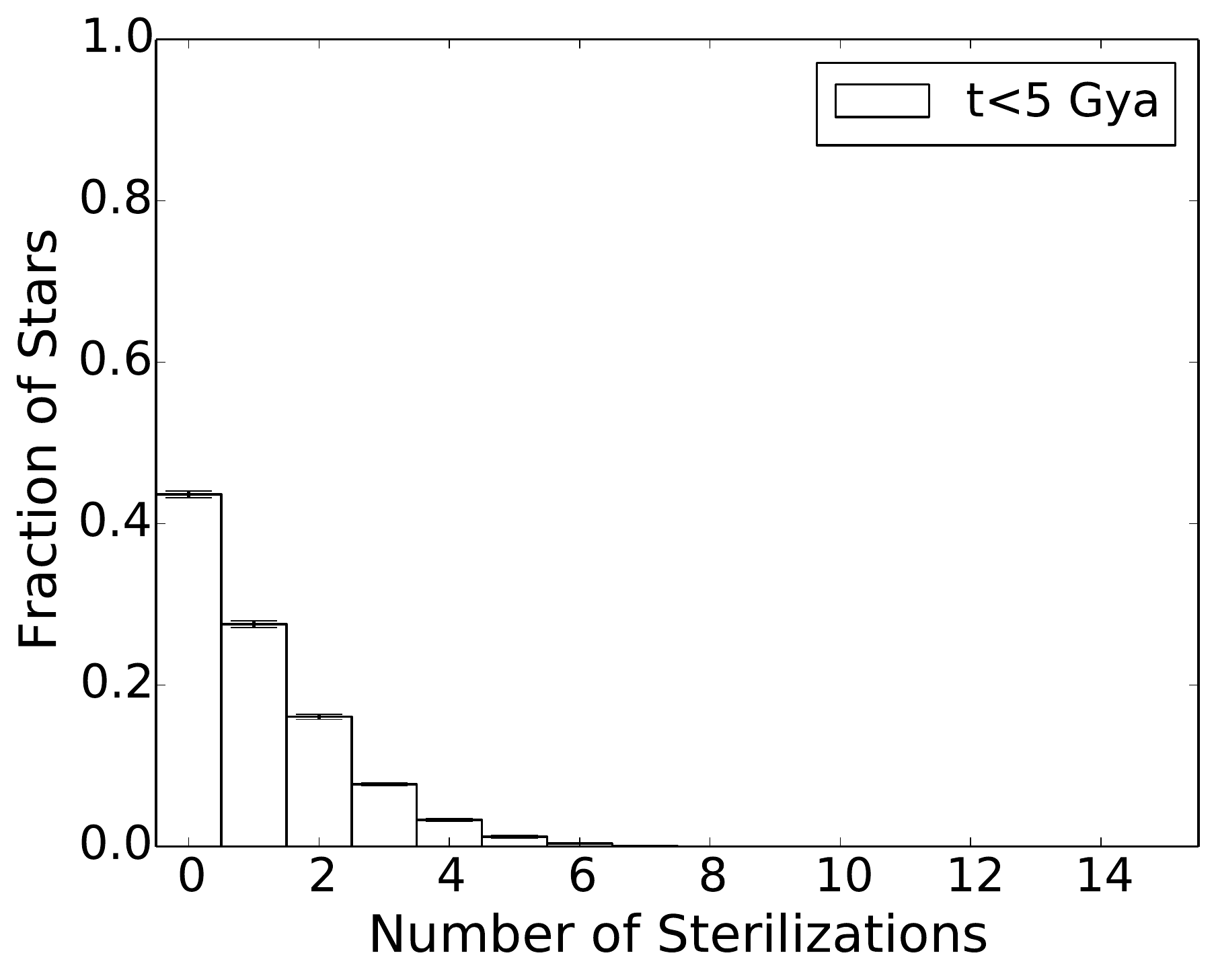}
      }

    \caption{The fraction of stars sterilized by the number of GRBs indicated on the horizontal axis over the past 1 and 5 Gyr at 8 kpc (a) and (b) and 14 kpc (c) and (d), respectively. Error bars represent 90\% confidence intervals.}
   \label{fig:last_1Gyr_5_gyr_8_14kpc_Model2}
\end{figure*}

Figure~\ref{fig:last_1Gyr_5_gyr_8_14kpc_Model2} shows the distribution of the number of sterilizations at $t<1$ Gya and $t<5$  Gya at two regions, $R=8$ kpc and $R=14$ kpc.  At $R=8$ kpc over the past 1 Gyr (5 Gyr), there is a $\sim$90\% ($\sim$30\%) chance of surviving a GRB.  Thus, assuming that GRBs occur in low metallicity environments, the solar neighborhood receives a significantly lower exposure rate to GRBs than if we assume they occur purely as a function of the SFH.  However, recent epochs in the history of the Galaxy are less favorable to the outskirts, where we find that over the past 1 Gyr $\sim$30\% of stars will be exposed to a GRB (Figure~\ref{fig:last_1Gyr_5_gyr_8_14kpc_Model2}~(c)).  However, this result assumes that the present day GRB rate, $\rho_{MW}$, is applicable when there is a metallicity dependence on GRB progenitors. As discussed in Section~\ref{method:GRB}, Model~2 may overestimate the lethality of GRBs, thus this may be considered an upper limit to the lethality of GRBs in the  outskirts.

\subsubsection{Surface Density of Surviving Stars}\label{results:model2_number}

\begin{figure*}[htp]
\centering
            \includegraphics[width=0.5\textwidth]{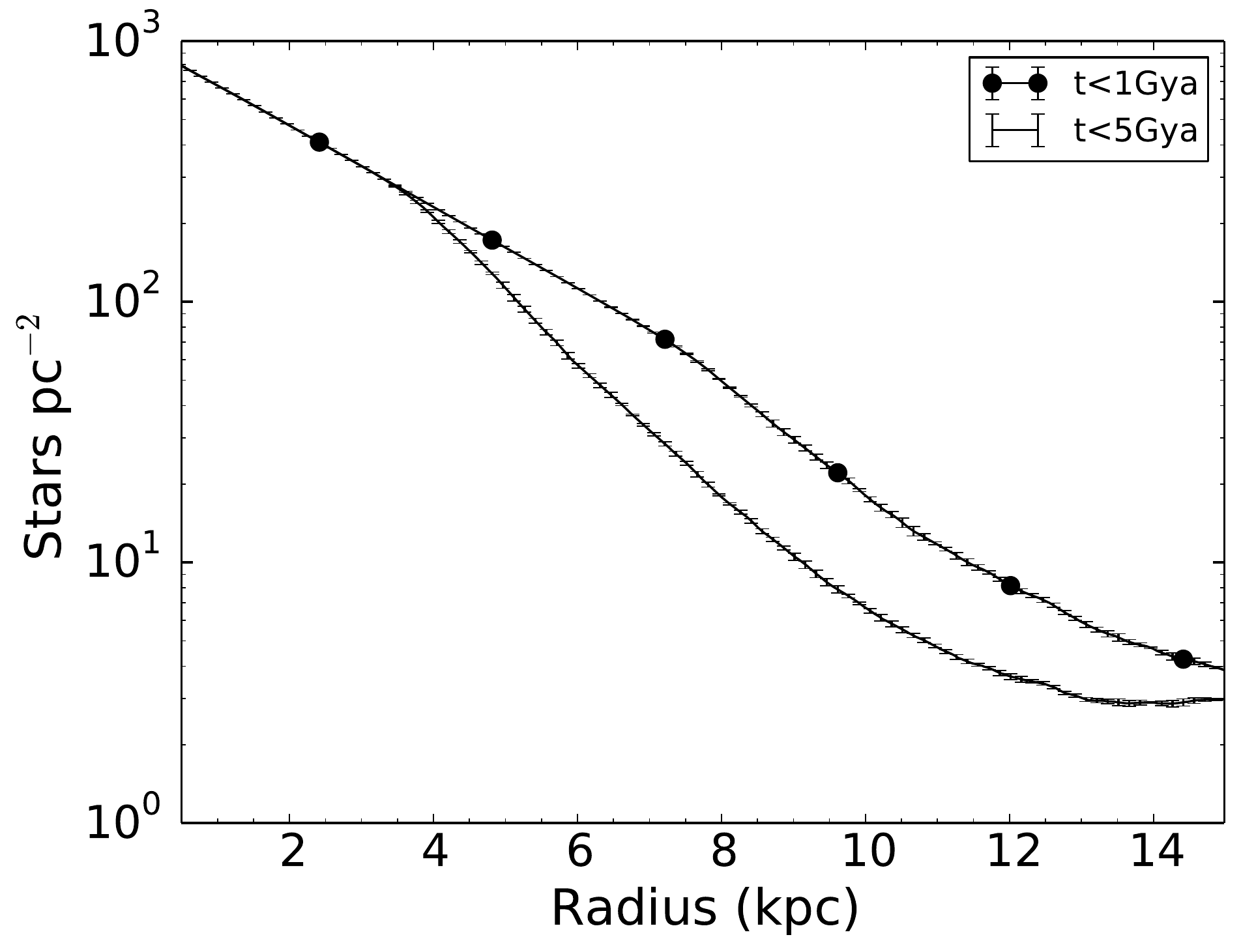}   
    \caption{The surface density of stars that are not within the beam of a GRB over the past 1 Gyr (upper curve) and 5 Gyr (lower curve). Error bars represent 90\% confidence intervals.}
   \label{fig:starsunsterilizedlastonegyrfivegyr-Model2}
\end{figure*}

The area density of stars that survive a GRB over the past 1 and 5 Gyr in Model~2 is plotted in Figure~\ref{fig:starsunsterilizedlastonegyrfivegyr-Model2}.  In recent epochs we observed that stars in the inner Galaxy all survive GRBs (Figure~\ref{fig:model2_grb}). This explains the identical number density of surviving stars over the past 1 and 5 Gyr at $R\lesssim 3$ kpc in Figure~\ref{fig:starsunsterilizedlastonegyrfivegyr-Model2}. Over the past 1 Gyr, we find that the surface density of surviving stars at $R=2.5$ kpc is $\sim400$ stars pc$^{-2}$ whereas $\sim50$ stars pc$^{-2}$ survive at $R=8$ kpc.  Since the inner Galaxy is devoid of GRBs within the last few Gyr due to sufficient chemical evolution, we find the inner Galaxy at recent epochs to be $\sim8\times$ more habitable than the position of the Earth at $R_\odot=8$ kpc.   The region of the Galaxy that has the greatest luminosity (the inner Galaxy) has the lowest chance of producing GRBs within the last 5 Gyr.  This is an interesting result as the inner Galaxy also contains the majority of the planets in the Milky Way, and our findings imply that the habitability of these planets will rarely be disrupted by GRB events.



\subsection{Discussion}\label{sec:comparison}
We contrast two scenarios regarding the metallicity evolution of the Milky Way, where one model populates GRBs purely as a function of the SFH (Model~1), and the other model includes a metallicity dependence on GRB formation (Model~2).  In Model~1 we find that at $R=8$ kpc, over the past 1 Gyr $\sim 35$\% of stars are exposed to a GRB, whereas \citet{2014PhRvL.113w1102P} find that this value is $60$\%.  This discrepancy in our results may be because we consider both the 3-dimensional stellar number density distribution and derive the GRB rate differently. Furthermore, the parameters in Equation~\ref{eqn:grbrate} may underestimate the absolute GRB rate at $z=0$. 


There has been speculation that GRBs will be prohibitive to life at high redshift due to both the increased SFR (which long GRBs are expected to trace), and lower metallicities, which are expected to increase the GRB formation rate. Although this work is focused on recent epochs, \citet{2015ApJ...810...41L} addressed whether life can survive at $z>0.5$ in the Milky Way by examining the dependence of the SFR and metallicity on the GRB formation rate.  They find that life can survive at high redshifts, assuming that life can survive a GRB every 0.5 Gyr. Our work broadly agrees with their findings, although our methods differ.  The model proposed in~\citet{2015ApJ...810...41L} does not consider chemical evolution, and is similar to Model~1 that favors populating GRBs in regions of the Milky Way with high luminosity, and not low metallicity.  Comparing~\citet{2015ApJ...810...41L} to Model~1, we also find that many planets in the Milky Way are likely to survive GRB events at $z\sim0.5$ ($\sim 5$ Gya). We find that at higher redshifts i.e., $z>0.5$, a large fraction of planets in the Milky Way survive lethal GRB events, with the exception of the first few Gyr.

There have been interesting recent developments regarding long GRBs.  \cite{2015ApJS..218...13Y} found that the local GRB formation rate at $z<1.0$ is roughly constant which is different than the SFR, whereas at $z>1.0$ the GRB rate is consistent with the SFR.  Assuming that this holds true for a small volume like the Milky Way, then we would expect that the metallicity dependence on the GRB formation rate would be less important at higher redshifts.  However, it is unclear as to whether the results of studies on the scale of cosmological volumes can be directly applied to smaller volumes such as the Galaxy.  \cite{2015ApJ...806...44P} finds that from the sample of GRBs used in their study, the GRB formation rate is expected to be lower than they found for low redshift, high metallicity galaxies.  Reconciling the differences between low and high redshift GRBs and metallicity effects on the GRB formation history require separate studies to establish the consequences of these findings for the habitability of the Galaxy.

Numerous works have examined the habitability of the Milky Way~\citep{2004Sci...303...59L,2011AsBio..11..855G,2015AsBio..15..683M,IJA:10152160,Vukotic2016} and have considered the effects of supernovae on planets. The studies vary in terms of where they find the most habitable region of the Galaxy at the present day. However, \cite{Vukotic2016} find the outskirts to be the most habitable region. Assuming GRBs favor low metallicity environments, from Model~2 we find that they may considerably reduce habitability at that location over the past $\sim5$ Gyr.

There are a number of uncertainties in the models.  Although we know that the GRB formation rate is a function of both the SFH and metallicity evolution, it is not entirely understood how these factors influence the GRB formation rate over time in the Milky Way.  Although our model produces more GRBs in the early history of the Galaxy with a rate that expectedly declines over time (Figure~\ref{fig:GRB_rates_model1_vs_model2}), our assumptions may under- or overestimate the GRB formation rate over the history of the Milky Way.  Furthermore, while the GRB rate has been studied in the literature in the context of a cosmological volume, there are uncertainties when applying this rate to a smaller volume such as the Milky Way even without accounting for the metallicity bias.  The present day GRB rate was derived and utilized in both models.  For consistency with Model~1 and to reduce the likelihood of underestimating the GRB rate, we elected to normalize to $\rho_{MW}$ in the metallicity dependent progenitor model (Model~2). As noted in Section~\ref{method:GRB}, another logically consistent method for producing GRBs in Model~2 would be to use the mass normalization in Model~1, which would reduce the present day GRB rate in Model~2.  We note that we have not modeled the bulge, and that this stellar population will decrease the overall propensity for life in the inner Galaxy. While we are aware of this uncertainty, we expect this danger to not significantly contribute to the population of GRBs within the last $\sim$5 Gyr, as the bulge contains an older stellar population that has undergone significant chemical evolution. Therefore, most of the GRBs in this region would have occurred at earlier epochs, and the metallicity enrichment may diminish GRB formation in this region, as was found for the disk (Figure~\ref{fig:model2_grb}). Additionally, the biological effects of a transient radiation event such as a GRB are not entirely known. Recently~\cite{2015AsBio..15..207T} found that the effects of UV radiation on life after ozone depletion may be lower than previous estimates.  Overall, GRBs may be both uncommon and less damaging than expected.

\section{CONCLUSIONS}\label{sec:conclusions}
We examine the lethality of GRBs in the Milky Way by modeling the 3-dimensional stellar number density distribution, inside-out formation history, and chemical evolution of the Galaxy.  We have proposed two models of the GRB formation history that differ as a function of the metallicity required of a GRB progenitor.  When we compare the two models, we find that when the GRB rate is a function of the SFH, the rate is roughly consistent over time, favoring GRBs that are located in the inner galaxy over all epochs. Whereas if we include the constraint that assumes GRBs can only form in low metallicity environments, the outskirts primarily hosts GRBs at recent epochs.  Due to declining stellar density with increasing galactocentric radius, low redshift GRBs located at the outskirts sterilize on average far fewer planets than GRBs at lower radii. Therefore the metallicity dependent progenitor model suggests that GRBs do not pose a significant danger to biospheres in the Milky Way at the present day.  Even if the GRB formation rate only follows the SFH, we still find that over the past 1 Gyr, the inner Galaxy hosts the greatest density of stars (and associated planets) that survive GRB events.  Thus, the region of the Milky Way with the greatest luminosity is the most favorable for life.  These calculations suggest that GRBs may be less lethal than previous estimates~\citep{1995ApJ...444L..53T,2014PhRvL.113w1102P,2015ApJ...810...41L}. Additionally, studies of progenitor environments suggest that GRBs are unlikely in the Milky Way~\citep{2006Natur.441..463F}, and GRBs have only been found in galaxies having $<10^{10}$M$_{\odot}$~\citep{2013ApJ...773..126J}, which excludes the Milky Way.  The results in this work, and others found in the literature imply that GRBs may be uncommon in the Milky Way and may not pose a significant danger to the propensity of planets to host life in the Galaxy.

\section*{ACKNOWLEDGEMENTS}
This paper has benefited from a thorough review of the manuscript by Brian Thomas and the insightful comments and fruitful discussions with David Patton, Henri Casanova, Ian Morrison, Chris Tinney, and Cody Rude. I would particularly like to thank the anonymous reviewer for all of the assistance and helpful comments provided throughout the review process.  The author would like to acknowledge the use of the University of Hawai`i High Performance Computing facilities, and in particular, technical support from Sean Cleveland and David Schanzenbach.

\bibliography{bibliography}

\end{document}